\documentclass{aa}
\usepackage{graphicx}
\usepackage{txfonts}
\usepackage{ulem}

\usepackage[colorlinks=true,citecolor=blue]{hyperref}
\usepackage{natbib}
\begin{document}

   \title{Detection of Na in WASP-21b's lower and upper atmosphere}


   \author{G. Chen\inst{1}
          \and
          N. Casasayas-Barris\inst{2,3}
          \and
          E. Pall\'{e}\inst{2,3}
          \and
          L. Welbanks\inst{4}
          \and
          N. Madhusudhan\inst{4}
          \and
          R. Luque\inst{2,3}
          \and
          F. Murgas\inst{2,3}
          }

   \institute{Key Laboratory of Planetary Sciences, Purple Mountain Observatory, Chinese Academy of Sciences, Nanjing 210023, PR China\\
         \email{guochen@pmo.ac.cn}
         \and
         Instituto de Astrof\'{i}sica de Canarias, V\'{i}a L\'{a}ctea s/n, E-38205 La Laguna, Tenerife, Spain
         \and
             Departamento de Astrof\'{i}sica, Universidad de La Laguna, Spain
         \and
             Institute of Astronomy, University of Cambridge, Madingley Road, Cambridge CB3 0HA, UK
             }

   \date{Received Month 00, 2020; accepted Month 00, 2020}

 
  \abstract
  {Optical transmission spectroscopy provides crucial constraints on the reference pressure levels and scattering properties for hot Jupiter atmospheres. For certain planets, where alkali atoms are detected in the atmosphere, their line profiles could serve as a good probe to link upper and lower atmospheric layers. WASP-21b is a Saturn-mass hot Jupiter orbiting a thick disc star, with a low density and an equilibrium temperature of 1333 K, which makes it a good target for transmission spectroscopy. Here, we present a low-resolution transmission spectrum for WASP-21b based in one transit observed by the OSIRIS spectrograph at the 10.4 m Gran Telescopio Canarias (GTC), and a high-resolution transmission spectrum based in three transits observed by HARPS-N at Telescopio Nazinale Galileo (TNG) and HARPS at the ESO 3.6 m telescope. We performed spectral retrieval analysis on GTC's low-resolution transmission spectrum and report the detection of Na at a confidence level of $>$3.5-$\sigma$. The Na line exhibits a broad line profile that can be attributed to pressure broadening, indicating a mostly clear planetary atmosphere. The spectrum shows a tentative excess absorption at the K D$_1$ line. Using HARPS-N and HARPS, we spectrally resolved the Na doublet transmission spectrum. An excess absorption at the Na doublet is detected during the transit, and shows a radial velocity shift consistent with the planet orbital motion. We proposed a metric to quantitatively distinguish hot Jupiters with relatively clear atmospheres from others, and WASP-21b has the largest metric value among all the characterized hot Jupiters. The detection of Na at both lower and upper atmosphere of WASP-21b reveals that it is an ideal target for future follow-up observations, providing the opportunity to understand the nature of its atmosphere across a wide range of pressure levels. }


   \maketitle
%

\section{Introduction}
\label{sec:intro}

Transmission spectroscopy \citep{2000ApJ...537..916S,2001ApJ...553.1006B} is one of the most efficient techniques to characterize exoplanet atmospheres. The slant viewing geometry makes it extremely sensitive to opacity sources in the atmospheres \citep{2005MNRAS.364..649F}, resulting in detections of a variety of atoms, ions, and molecules in dozens of exoplanets \citep[e.g.,][]{2016Natur.529...59S,2018AJ....155..156T,2019ARA&A..57..617M}. Consequently, trends start to emerge in the derived chemical abundances and metallicities, which could connect to planet formation histories \citep[e.g.,][]{2014ApJ...793L..27K,2014ApJ...791L...9M,2017MNRAS.469.4102M,2017Sci...356..628W,2019MNRAS.482.1485P,2019ApJ...887L..20W}. 

However, given the degeneracy between reference pressure and chemical abundances \citep{2012ApJ...753..100B,2014RSPTA.37230086G,2017MNRAS.470.2972H}, ubiquitous clouds and hazes strongly degrade our capability to precisely retrieve detailed atmospheric nature \citep{2016ApJ...817L..16S,2016ApJ...823..109I,2016ApJ...826L..16H,2017AJ....154..261C,2017ApJ...847L..22F}. One way to break the degeneracy is to search for the pressure broadening signature of alkali lines \citep{2014RSPTA.37230086G,2017MNRAS.470.2972H,2017MNRAS.469.1979M,2019AJ....157..206W}, in particular Na and K, which, when resolved at high spectral resolution, could also help characterize planetary wind and give insight into heating and cooling processes in the upper atmosphere \citep{2015ApJ...814L..24L,2017ApJ...851..150H,2020A&A...633A..86S,2020arXiv200502536G}. Recently, several hot Jupiters have been found to exhibit broad line profiles, at the Na or K lines, that can be associated with pressure broadening \citep[e.g.,][]{2018Natur.557..526N,2018A&A...616A.145C,2019AJ....157...21P}. All these planets seem to have equilibrium temperatures clustered between 1200~K and 1500~K.

Here we present low- and high-resolution transit observations of the Saturn-mass hot Jupiter WASP-21b. This low-density planet has an equilibrium temperature of $T_\mathrm{eq}=1333\pm28$~K and a low surface gravity of $g_\mathrm{p}=5.07\pm0.35$~m\,s$^{-1}$ \citep{2013A&A...557A..30C}, which could potentially exhibit a transit depth variation of 251~ppm per scale height, making it a good target for atmospheric characterization via transmission spectroscopy. WASP-21b orbits a G3V thick disc star in a circular orbit every 4.32~days \citep{2010A&A...519A..98B}, which is one of the most metal-poor planet hosts ($\mathrm{[Fe/H]}=-0.46\pm0.11$). \citet{2011MNRAS.416.2593B} analyzed three transits, two of which are partial transits, obtained with the robotic 2.0~m Liverpool Telescope, and found that the host star is evolving off the main sequence. They revised down the stellar mass and hence obtained a lower planet mass. \citet{2012MNRAS.426.1291S} reanalyzed the data of \citet{2011MNRAS.416.2593B} in the homogeneous studies of 38 planets, and derived a stellar mass closer to \citet{2010A&A...519A..98B} but a stellar radius larger than both \citet{2010A&A...519A..98B} and \citet{2011MNRAS.416.2593B}. \citet{2013A&A...557A..30C} observed a new single transit with both 1.5~m Cassini Telescope and 1.2~m Calar Alto Telescope, and presented the latest revised physical parameters. They derived $0.890\pm0.079$~$M_\sun$ and $1.136\pm0.051$~$R_\sun$ for the host star, and $0.276\pm0.019$~$M_\mathrm{jup}$ and $1.162\pm0.054$~$R_\mathrm{jup}$ for the planet. Finally, \citet{2015MNRAS.451.4060S} added another three new transit observations and found no hints of significant transit timing variations.

This paper is organized as follows. In Sect.~\ref{sec:obs}, we summarize the low- and high-resolution transit observations and detail the data reduction. In Sect.~\ref{sec:lowres_analysis}, we present the light-curve analysis for the low-resolution data, and describe the spectral retrieval analysis. In Sect.~\ref{sec:hires_analysis}, we present the analyses on radial velocities and high-resolution transmission spectroscopy. In Sect.~\ref{sec:discuss}, we discuss the properties of WASP-21b's atmosphere inferred from transmission spectrum, and put it in the context of all hot Jupiters that have been characterized by low-resolution optical transmission spectroscopy. Conclusions are given in Sect.~\ref{sec:conclusions}.

\section{Observations and data reduction}
\label{sec:obs}

\begin{table*}
     \centering
     \caption{Observation summary.}
     \label{tab:obslog}
     \begin{tabular}{ccccccccccc}
     \hline\hline\noalign{\smallskip}
     \# & Telescope & Instrument & Start Night & UT window & $T_\mathrm{exp}$ & $N_\mathrm{obs}$ & Airmass\tablefootmark{(a)} & \multicolumn{2}{c}{$S/N$\tablefootmark{(b)}} & Program\\\noalign{\smallskip}
     \cline{9-10}\noalign{\smallskip}
      &  &  &  &  & [s] &  &  & Cont. & Core & \\\noalign{\smallskip}
     \hline\noalign{\smallskip}
     1 & ESO 3.6~m & HARPS & 2011-09-05 & 02:37-08:18 & 900 & 22 & 1.94-1.48-2.33 & 29-37 & 8-11 & 087.C-0649(A) \\\noalign{\smallskip}
     2 & ESO 3.6~m & HARPS & 2011-09-18 & 01:44-06:54 & 900 & 19 & 1.96-1.48-1.98 & 21-29 & 6-9 & 087.C-0649(A) \\\noalign{\smallskip}
     3 & TNG & HARPS-N & 2018-09-07 & 21:37-06:00 & 900 & 33 & 1.58-1.02-2.15 & 23-45 & 7-13 & CAT18A\_D1 \\\noalign{\smallskip}
     \hline\noalign{\smallskip}
     4 & GTC & OSIRIS & 2012-09-11 & 20:41-00:48 & 22 & 313 & 1.93-1.02-1.02 & -- & -- & GTC47-12B \\\noalign{\smallskip}
     \hline\noalign{\smallskip}
    \end{tabular}
    \tablefoot{
     \tablefoottext{a}{The first and third values refer to the airmass at the beginning and at the end of the observation. The second value gives the minimum airmass.}
     \tablefoottext{b}{The two values correspond to the minimum and maximum signal-to-noise ratio (S/N), respectively. The S/N of the continuum was measured at around 5888~$\AA$. The S/N of the Na core was measured at the D$_2$ line.}
     }
\end{table*}

To derive the transmission spectrum for WASP-21b, we observed one transit at low spectral resolution and one transit at high spectral resolution. We also collected archival data for another two transits observed at high spectral resolution. The low-resolution observation was carried out in seeing-limited conditions, along with a reference star, while the high-resolution observations were single-object only, without flux calibration. The summary of the five transit observations is given in Table \ref{tab:obslog}. 

\subsection{GTC/OSIRIS}
\label{sec:gtcobs}

One transit of \object{WASP-21b} was observed on the night of September 11, 2012 (program GTC47-12B, PI: E.~Pall\'{e}), using the OSIRIS spectrograph \citep{2012SPIE.8446E..4TS} installed at the 10.4~m Gran Telescopio CANARIAS (GTC) in La Palma, Spain. The observation was performed with the R1000R grism through the 12$''$ slit. The R1000R grism can cover a wavelength range of 510--1000~nm at a spectral resolution of $\mathcal{R}\sim 1122$. The long-slit allows a reference star to be simultaneously observed with the target star \object{WASP-21} ($r'=11.4$~mag). The adopted reference star \object{2MASS J23094822+1822564} ($r'=11.6$~mag) was 2.5$'$ away. Both stars were placed on CCD chip 2, while CCD chip 1 was switched off. The CCD was configured in the $2\times$2 binning mode (0.254$''$ per binned pixel) with a readout speed of 200~kHz. 

The observation lasted 4.1 hours, and missed the pre-transit baseline while included 42~min of post-transit baseline. The first eleven frames had an exposure time of 30 sec, while the remaining used 22 sec. A total of 313 frames were recorded. The resulting duty cycle is 47.1\%. The weather was not clear all the time. The stars might have passed thin cirrus for $\sim$1.1~hours after mid-transit, during which the target and reference stars showed similar flux variation. During the whole observation, the airmass dropped monotonically from 1.928 to 1.017. The seeing varied between 0.73$''$ and 1.78$''$, which was measured as the full width at half maximum (FWHM) of the spatial profile at the central wavelength. This resulted in a seeing-limited spectral resolution of roughly 10~$\AA$. The centroids of the spatial profile drifted around 2 pixels, while no clear drift trend was observed in the cross-dispersion direction. 

\begin{figure}
\centering
\includegraphics[width=1\linewidth]{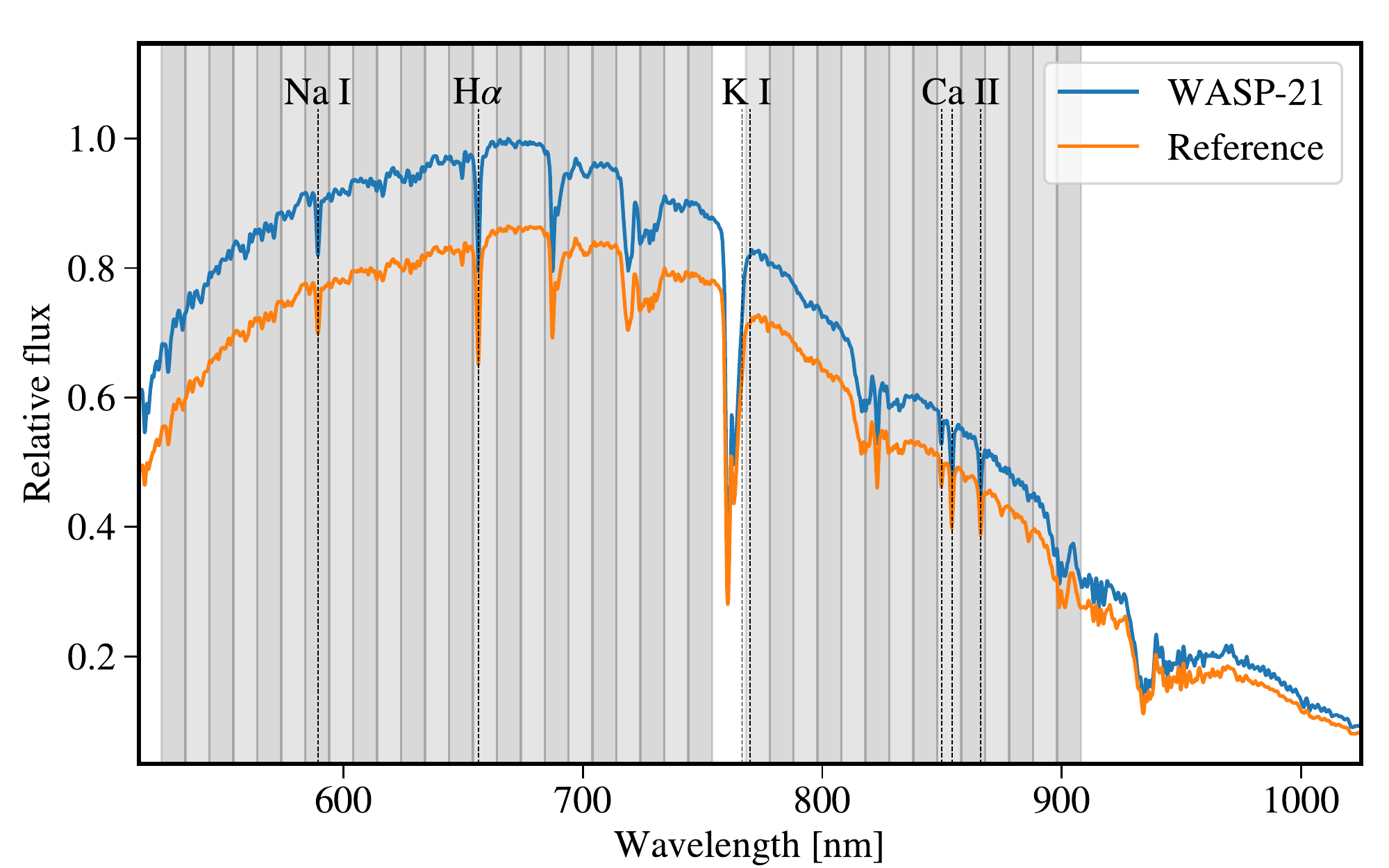}
\caption{Median-combined out-of-transit stellar spectra of \object{WASP-21} (blue) and the reference star (orange), obtained with the R1000R grism of GTC/OSIRIS on the night of September 11, 2012. The spectra have been individually normalized, with the target-to-reference flux ratio being preserved. The color-shaded areas indicate the divided passbands that are used to create the spectroscopic light curves.}
\label{fig:GTCSpectra}
\end{figure}

The spectral images were calibrated following the same method adopted in \citet{2017A&A...600A.138C,2017A&A...600L..11C,2018A&A...616A.145C}, including overscan and bias subtraction, flat correction, and sky removal. The one-dimensional spectra were extracted using an aperture diameter of 42~pixels, which gave the lowest scatter in the white-color light curve. The time stamp was converted to Barycentric Julian Date in the Barycentric Dynamical Time standard \citep[$\mathrm{BJD}_\mathrm{TDB}$;][]{2010PASP..122..935E}. The white-color light curve was integrated between 524~nm and 908~nm, except that 754--768~nm was excluded to avoid the strong noise introduced by the telluric oxygen-A band. The spectral light curves were integrated in 10~nm bin width. The wavelengths longer than 908~nm were not used owing to significant fringing effects. Figure~\ref{fig:GTCSpectra} shows the median-combined out-of-transit stellar spectra and the adopted spectral band passes.

\subsection{TNG/HARPS-N}

One transit of WASP-21b was observed on the night of September 7, 2018 (program CAT18A\_D1, PI: G.~Chen) using the HARPS-N spectrograph \citep{2014SPIE.9147E..8CC} installed at the 3.58~m Telescopio Nazionale {\it Galileo} (TNG) in La Palma, Spain. HARPS-N is a fiber-fed echelle spectrograph, covering a wavelength range of 383--693~nm at a spectral resolution of $\mathcal{R}\sim 115\,000$. One of the two HARPS-N fibers (fiber A) was pointed to the target star WASP-21, and the other (fiber B) was pointed to sky to monitor the telluric emission. The observation lasted 8.4 hours, and covered the whole transit. The exposure time was 900 sec, resulting in 33 frames, of which 11 frames were fully in-transit (i.e., between the second and third contacts of a transit) and 19 frames were fully out-of-transit (i.e., no overlap with the transit during the exposure). The signal-to-noise ratio (S/N) varied between 23 and 45 at the continuum around 5888~$\AA$, while it decreased to S/N$\sim$7--13 at the Na D$_2$ line core.

The data were reduced by the version 3.7 of the HARPS-N data reduction software. In the subsequent analysis, we will use the pipeline-reduced order-merged one-dimensional spectra, labeled as \texttt{s1d} by the pipeline. The wavelengths have been corrected for the barycentric Earth radial velocity, and have been resampled in a step of 0.01~$\AA$. Strong interstellar Na absorption can be seen in the WASP-21 spectra. Fortunately, the systemic radial velocity of WASP-21 is well away from zero \citep[$-$89.45~km\,s$^{-1}$,][also see Sect.~\ref{sec:rm_analysis}]{2010A&A...519A..98B}, shifting the stellar Na well separated from interstellar Na. The interstellar Na were masked in the subsequent analysis. The telluric Na emission was close to the interstellar Na, which was simultaneously masked. Therefore, we did not use fiber B to correct the telluric emission in fiber A.

\subsection{ESO 3.6~m/HARPS}

The archival data for two transits of WASP-21b were collected from the ESO archive under the program 087.C-0649(A) (PI: A.~Triaud). The two transits covered the complete transit event, which were observed on the nights of September 5, 2011 and September 18, 2011, respectively. The observations were made with the HARPS spectrograph \citep{2003Msngr.114...20M} installed at the ESO 3.6~m telescope in La Silla, Chile. As the design predecessor of HARPS-N, HARPS also has a spectral resolution of $\mathcal{R}\sim 115\,000$ and covers a wavelength range of 378--691~nm. For the two archival transits, only the data collected by the fiber A were available. An exposure time of 900~sec was adopted for the two transits. The first transit was observed for 5.7 hours, with 22 frames being collected, of which 11 were fully in-transit and 8 were fully out-of-transit. The second transit was observed for 5.2 hours, with 19 frames being collected, of which 11 were fully in-transit and 6 were fully out-of-transit. The S/N of the HARPS spectra were similar to those of the HARPS-N spectra, but with slightly lower values due to higher airmass visible from the Southern hemisphere. The data were reduced by the version 3.5 of the HARPS data reduction software. The pipeline products are similar to the HARPS-N ones.

\section{Low-resolution data analysis}
\label{sec:lowres_analysis}

\subsection{GTC/OSIRIS light-curves}
\label{sec:gtclc}

\begin{table}
     \centering
     \caption{Adopted parameters for the WASP-21 system.}
     \label{tab:adopted_param}
     \begin{tabular}{lr}
     \hline\hline\noalign{\smallskip}
     Parameter & Value \\\noalign{\smallskip}
     \hline\noalign{\smallskip}
     \noalign{\smallskip}\multicolumn{2}{c}{\it Stellar Parameters}\\\noalign{\smallskip}
       Stellar mass, $M_\star$ [$M_\sun$]              & $0.890 \pm 0.079$       \tablefootmark{~(a)}     \\
       Stellar radius, $R_\star$ [$R_\sun$]            & $1.136 \pm 0.051$       \tablefootmark{~(a)}     \\
       Effective temperature, $T_\mathrm{eff}$ [K]     & $5800 \pm 100$          \tablefootmark{~(b)}     \\
       Surface gravity, $\log g_\star$ [cgs]           & $4.277 \pm 0.026$       \tablefootmark{~(a)}     \\
       Metallicity, $\mathrm{[Fe/H]}$                  & $-0.46 \pm 0.11$        \tablefootmark{~(b)}     \\
       Proj. rotation velocity, $v\sin i_\star$ [$\mathrm{km\,s}^{-1}$]  & $1.5 \pm 0.6$      \tablefootmark{~(b)}     \\\noalign{\smallskip}
     \noalign{\smallskip}\multicolumn{2}{c}{\it Planet Parameters}\\\noalign{\smallskip}
        Planet mass, $M_p$ [$M_J$]                     & $0.276 \pm 0.019$       \tablefootmark{~(a)}     \\
        Planet radius, $R_p$ [$R_J$]                   & $1.162 \pm 0.054$       \tablefootmark{~(a)}     \\
        Equilibrium temperature, $T_\mathrm{eq}$ [K]   & $1333 \pm 28$           \tablefootmark{~(a)}     \\
        Surface gravity, $\log g_p$ [cgs]              & $2.71 \pm 0.03$         \tablefootmark{~(a)}     \\\noalign{\smallskip}
     \noalign{\smallskip}\multicolumn{2}{c}{\it Orbital Parameters}\\\noalign{\smallskip}
        Semi-major axis, $a$ [AU]                      & $0.0499 \pm 0.0015$     \tablefootmark{~(a)}     \\
        Eccentricity, $e$                              & 0 (fixed)               \tablefootmark{~(b)}     \\
        RV semi-amplitude, $K_\star$ [$\mathrm{km\,s}^{-1}$] & $0.0372 \pm 0.0011$    \tablefootmark{~(b)}     \\
    \hline\noalign{\smallskip}
    \end{tabular}
    \tablebib{
      \tablefoottext{a}{\citet{2013A&A...557A..30C}.}
      \tablefoottext{b}{\citet{2010A&A...519A..98B}.}
    }
\end{table}

\begin{figure}
\centering
\includegraphics[width=1\linewidth]{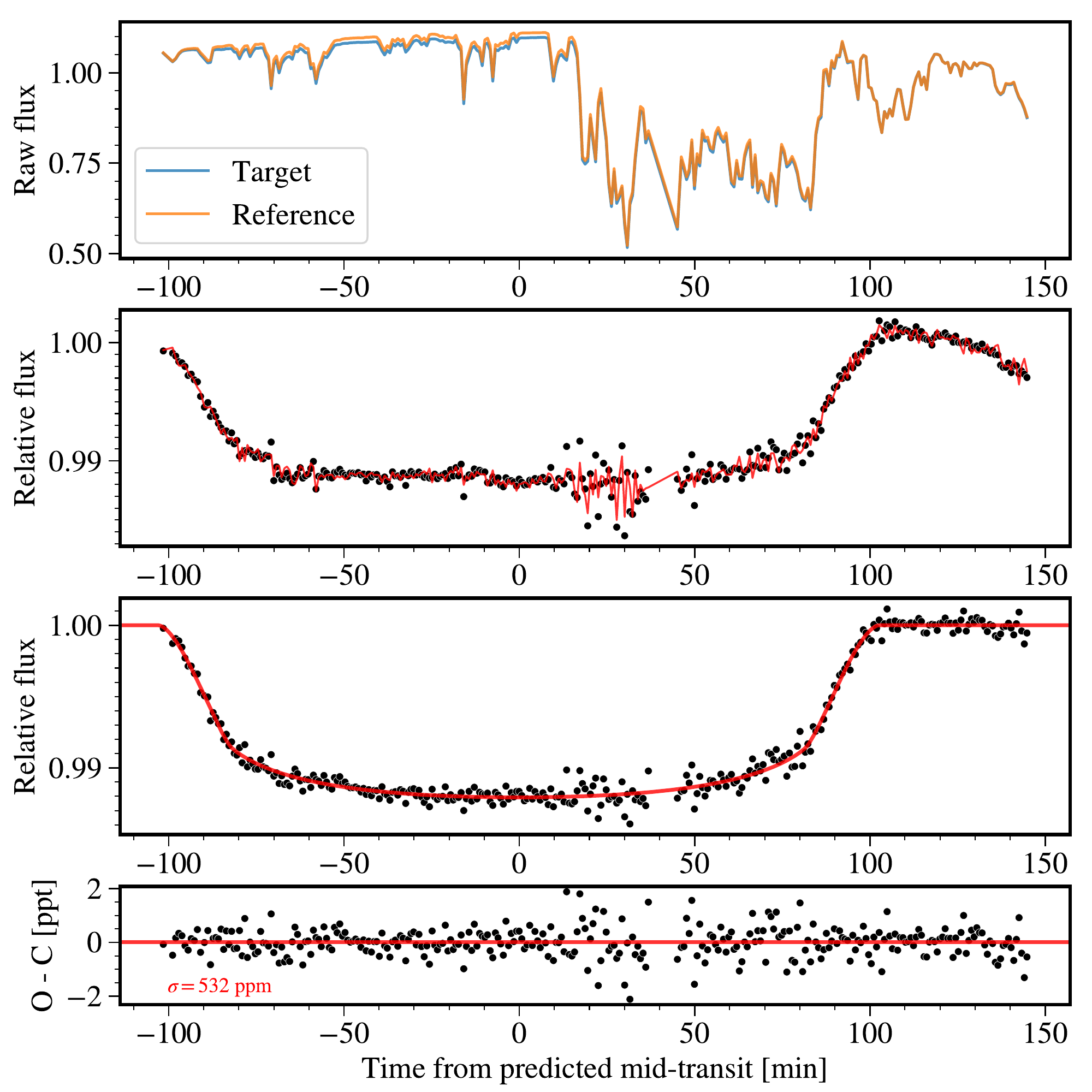}
\caption{White-color light curve obtained with GTC/OSIRIS. From top to bottom are {\it i)} raw flux time series divided by exposure time, {\it ii)} raw light curve (i.e., normalized target-to-reference flux ratios), {\it iii)} light curve corrected for systematics, and {\it vi)} best-fit residuals. The red line shows the best-fit model. The gap at around $+$40 min is due to missing raw files in the data archive.}
\label{fig:GTCWhiteLC}
\end{figure}

\begin{table}
     \centering
     \caption{Derived parameters from GTC/OSIRIS white light curve.}
     \label{tab:gtc_param}
     \begin{tabular}{lr}
     \hline\hline\noalign{\smallskip}
     Parameter & Value \\\noalign{\smallskip}
     \hline\noalign{\smallskip}
     \multicolumn{2}{c}{\it Free parameters}\\\noalign{\smallskip}
     Radius ratio, $R_p/R_\star$         & $0.1016 ^{+0.0015}_{-0.0014}$ \\\noalign{\smallskip}
     Orbital inclination, $i$ [deg]      & $88.28 ^{+0.90}_{-0.69}$ \\\noalign{\smallskip}
     Scaled semi-major axis, $a/R_\star$ & $10.23 ^{+0.36}_{-0.40}$ \\\noalign{\smallskip}
     Linear limb-darkening coeff., $u_1$ & $0.294 \pm 0.063$ \\\noalign{\smallskip}
     Quad. limb-darkening coeff., $u_2$  & $0.364 \pm 0.090$ \\\noalign{\smallskip}
     Mid-transit time , $T_{\rm mid}$ [MJD \tablefootmark{~(a)}] & $56181.93874 \pm 0.00031$ \\\noalign{\smallskip}
     \multicolumn{2}{c}{\it Derived parameters}\\\noalign{\smallskip}
     Transit duration, $T_{14}$ [d]                   & $0.1426 ^{+0.0013}_{-0.0012}$         \\\noalign{\smallskip}
     Fully in-transit duration, $T_{23}$ [d]          & $0.1137 ^{+0.0017}_{-0.0021}$         \\\noalign{\smallskip}
     \multicolumn{2}{c}{\it Revised ephemeris}\\\noalign{\smallskip}
     Transit epoch, $T_0$ [MJD \tablefootmark{~(a)}]  & $54742.54198 \pm 0.00068$ \\\noalign{\smallskip}
     Orbital period, $P$ [d]                          & $4.3225130 \pm 0.0000021$ \\\noalign{\smallskip}
    \hline\noalign{\smallskip}
    \end{tabular}
    \tablefoot{
      \tablefoottext{a}{$\mathrm{MJD}=\mathrm{BJD}_\mathrm{TDB}-2400000.5$.}
    }
\end{table}

\begin{table}
     \centering
     \caption{Derived GTC/OSIRIS transmission spectrum.}
     \label{tab:gtc_transpec}
     \begin{tabular}{cccc}
     \hline\hline\noalign{\smallskip}
\# & $\lambda_\mathrm{start}$ [nm] & $\lambda_\mathrm{end}$ [nm] & $R_\mathrm{p}/R_\star$\\\noalign{\smallskip}
     \hline\noalign{\smallskip}
 1 & 524 & 534 & $0.0996 ^{+0.0013 }_{-0.0012 }$\\ \noalign{\smallskip}
 2 & 534 & 544 & $0.1006 ^{+0.0012 }_{-0.0013 }$\\ \noalign{\smallskip}
 3 & 544 & 554 & $0.1003 ^{+0.0013 }_{-0.0013 }$\\ \noalign{\smallskip}
 4 & 554 & 564 & $0.1035 ^{+0.0013 }_{-0.0012 }$\\ \noalign{\smallskip}
 5 & 564 & 574 & $0.1033 ^{+0.0011 }_{-0.0011 }$\\ \noalign{\smallskip}
 6 & 574 & 584 & $0.1049 ^{+0.0011 }_{-0.0011 }$\\ \noalign{\smallskip}
 7 & 584 & 594 & $0.1087 ^{+0.0017 }_{-0.0017 }$\\ \noalign{\smallskip}
 8 & 594 & 604 & $0.1051 ^{+0.0011 }_{-0.0012 }$\\ \noalign{\smallskip}
 9 & 604 & 614 & $0.1049 ^{+0.0011 }_{-0.0011 }$\\ \noalign{\smallskip}
10 & 614 & 624 & $0.1024 ^{+0.0013 }_{-0.0014 }$\\ \noalign{\smallskip}
11 & 624 & 634 & $0.1052 ^{+0.0011 }_{-0.0011 }$\\ \noalign{\smallskip}
12 & 634 & 644 & $0.1043 ^{+0.0011 }_{-0.0012 }$\\ \noalign{\smallskip}
13 & 644 & 654 & $0.1026 ^{+0.0011 }_{-0.0011 }$\\ \noalign{\smallskip}
14 & 654 & 664 & $0.1026 ^{+0.0011 }_{-0.0012 }$\\ \noalign{\smallskip}
15 & 664 & 674 & $0.1025 ^{+0.0012 }_{-0.0012 }$\\ \noalign{\smallskip}
16 & 674 & 684 & $0.1003 ^{+0.0012 }_{-0.0012 }$\\ \noalign{\smallskip}
17 & 684 & 694 & $0.1036 ^{+0.0011 }_{-0.0011 }$\\ \noalign{\smallskip}
18 & 694 & 704 & $0.1003 ^{+0.0012 }_{-0.0013 }$\\ \noalign{\smallskip}
19 & 704 & 714 & $0.1029 ^{+0.0012 }_{-0.0012 }$\\ \noalign{\smallskip}
20 & 714 & 724 & $0.1012 ^{+0.0013 }_{-0.0013 }$\\ \noalign{\smallskip}
21 & 724 & 734 & $0.1004 ^{+0.0013 }_{-0.0012 }$\\ \noalign{\smallskip}
22 & 734 & 744 & $0.1022 ^{+0.0012 }_{-0.0012 }$\\ \noalign{\smallskip}
23 & 744 & 754 & $0.1017 ^{+0.0013 }_{-0.0014 }$\\ \noalign{\smallskip}
24 & 768 & 778 & $0.1027 ^{+0.0013 }_{-0.0013 }$\\ \noalign{\smallskip}
25 & 778 & 788 & $0.0998 ^{+0.0012 }_{-0.0013 }$\\ \noalign{\smallskip}
26 & 788 & 798 & $0.1022 ^{+0.0012 }_{-0.0012 }$\\ \noalign{\smallskip}
27 & 798 & 808 & $0.1019 ^{+0.0013 }_{-0.0012 }$\\ \noalign{\smallskip}
28 & 808 & 818 & $0.1008 ^{+0.0017 }_{-0.0015 }$\\ \noalign{\smallskip}
29 & 818 & 828 & $0.0995 ^{+0.0014 }_{-0.0014 }$\\ \noalign{\smallskip}
30 & 828 & 838 & $0.1047 ^{+0.0014 }_{-0.0013 }$\\ \noalign{\smallskip}
31 & 838 & 848 & $0.1022 ^{+0.0014 }_{-0.0016 }$\\ \noalign{\smallskip}
32 & 848 & 858 & $0.1014 ^{+0.0016 }_{-0.0017 }$\\ \noalign{\smallskip}
33 & 858 & 868 & $0.1007 ^{+0.0017 }_{-0.0017 }$\\ \noalign{\smallskip}
34 & 868 & 878 & $0.1004 ^{+0.0017 }_{-0.0017 }$\\ \noalign{\smallskip}
35 & 878 & 888 & $0.0989 ^{+0.0023 }_{-0.0022 }$\\ \noalign{\smallskip}
36 & 888 & 898 & $0.1000 ^{+0.0022 }_{-0.0022 }$\\ \noalign{\smallskip}
37 & 898 & 908 & $0.1013 ^{+0.0034 }_{-0.0035 }$\\ \noalign{\smallskip}
    \hline\noalign{\smallskip}
    \end{tabular}
\end{table}

\begin{figure*}
\centering
\includegraphics[width=\linewidth]{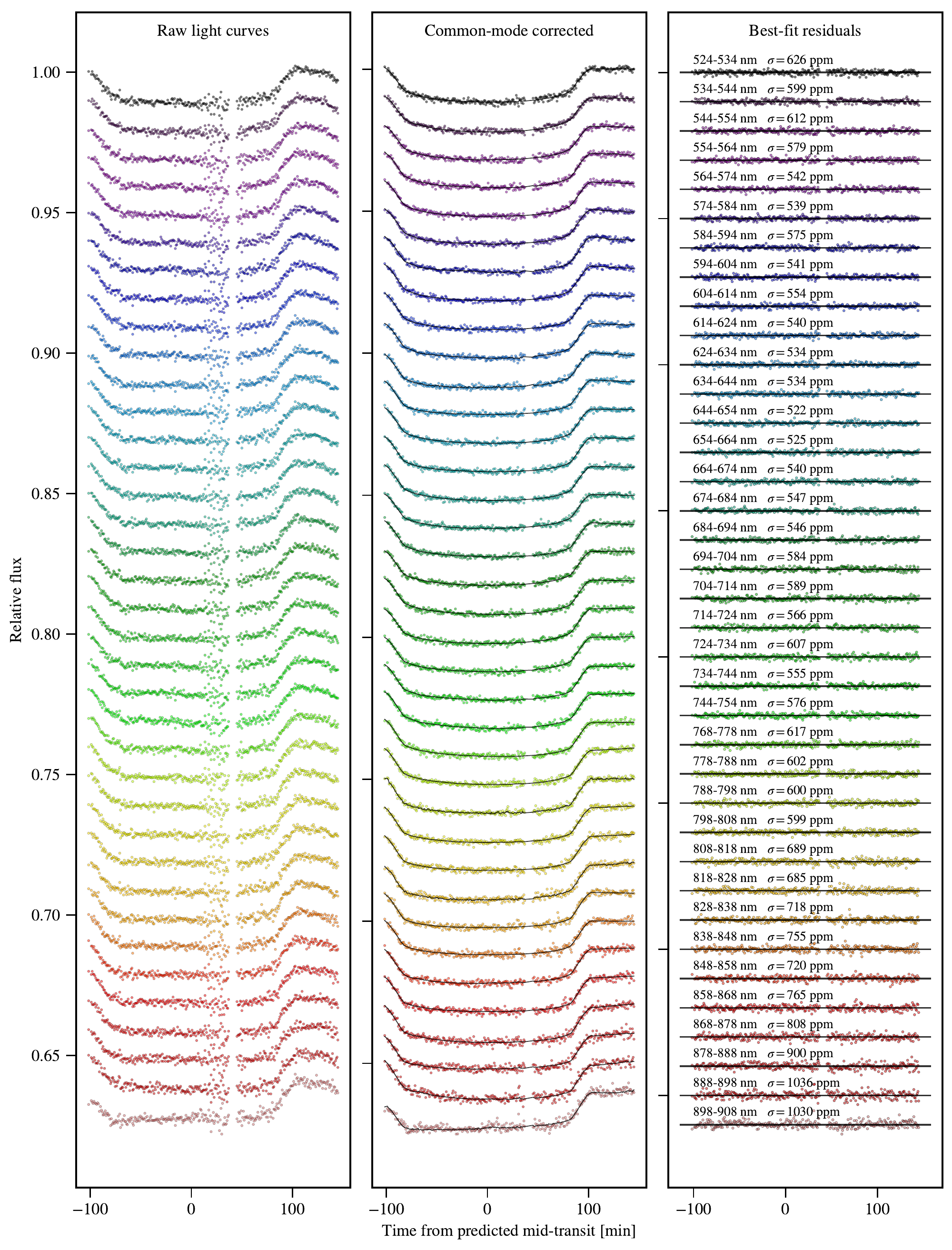}
\caption{Spectroscopic light curves of \object{WASP-21b} obtained with the R1000R grism of GTC/OSIRIS before ({\it left panel}) and after removing the common-mode systematics ({\it middle panel}), along with the best-fit residuals ({\it right panel}). }
\label{fig:GTCSpecLC}
\end{figure*}

The GTC/OSIRIS transit light curves were modeled following the method described in \citet{2018A&A...616A.145C}, where Gaussian processes \citep[GP;][]{2006gpml.book.....R,2012MNRAS.419.2683G} were employed to account for the correlated noise. The \citet{2002ApJ...580L.171M} analytic transit model and GP were implemented by the Python packages \texttt{batman} \citep{2015PASP..127.1161K} and \texttt{george} \citep{2015ITPAM..38..252A}, respectively. The Bayesian parameter estimation was implemented by the Python package \texttt{emcee} \citep{2013PASP..125..306F}.

We adopted the quadratic limb darkening law for the analytic transit model. The limb darkening coefficients ($u_1$, $u_2$) were fitted with Gaussian priors $\mathcal{N}(u_i,\sigma_{u_i}^2)$ for both white and spectral light curves. The prior mean values ($u_i$) were derived using the Kurucz ATLAS9 stellar atmosphere models with stellar effective temperature $T_\mathrm{eff}=5750$~K, surface gravity $\log g_\star=4.5$, and metallicity $\mathrm{[Fe/H]}=-0.5$ \citep{2015MNRAS.450.1879E}, which is the closest grid to the stellar parameters of \object{WASP-21} \citep[][also see our Table \ref{tab:adopted_param}]{2010A&A...519A..98B}. The prior standard deviation values were $\sigma_{u_i}=0.1$ for the white light curve, and the width of the model grid gap for the spectral light curves. The grid gap was estimated by comparing the set of $T_\mathrm{eff}=5750$~K to another two sets of values calculated at $T_\mathrm{eff}=5500$~K and $T_\mathrm{eff}=6000$~K, where the larger difference was recorded. 

We employed the squared exponential (SE) kernel for the GP covariance matrix,
\begin{align}
k_\mathrm{SE}(x_i,x_j,\theta)&=A^2\exp\Bigg[-\sum\limits_{\alpha=1}^{N}\bigg(\frac{x_{\alpha,i}-x_{\alpha,j}}{L_\alpha}\bigg)^2\Bigg]. 
\end{align}
For the white-color light curve, we used the analytic transit model $\mathcal{T}_\mathrm{w}$ as the mean function of the GP, with inclination $i$, scale semi-major axis $a/R_\star$, radius ratio $R_\mathrm{p}/R_\star$, mid-transit time $T_\mathrm{mid}$, u$_1$, and u$_2$ as the free parameters. Corresponding GP input variables $x_{i,j}$ were time sequence $t$, spatial position drift $y$ and spatial seeing variation $s_y$. After the light-curve modeling, we  divided the white-color light curve by the best-fitting transit model $\mathcal{T}_\mathrm{w}$ to derive the common-mode trend $\mathcal{S}_\mathrm{w}$, and then divided every spectral light curve by $\mathcal{S}_\mathrm{w}$ to correct for this common-mode trend. To model the corrected spectral light curves, we used the analytic transit model multiplied by a baseline function $\mathcal{T}_\mathrm{spec}(c_0+c_1t+c_2t^2)$ as the mean function. The free transit parameters were $R_\mathrm{p}/R_\star$, $u_1$, and $u_2$, while the others were fixed to the white light-curve values (see Table~\ref{tab:gtc_param}). In this case, the GP input variables $x_{i,j}$ were $y$ and $s_y$. All the GP hyperparameters were fit with uniform priors. The time scale hyperparameter $L_t$ is always forced to be no shorter than \object{WASP-21b}'s ingress/egress duration (0.01445 days). 

\begin{figure*}
\centering
\includegraphics[width=\linewidth]{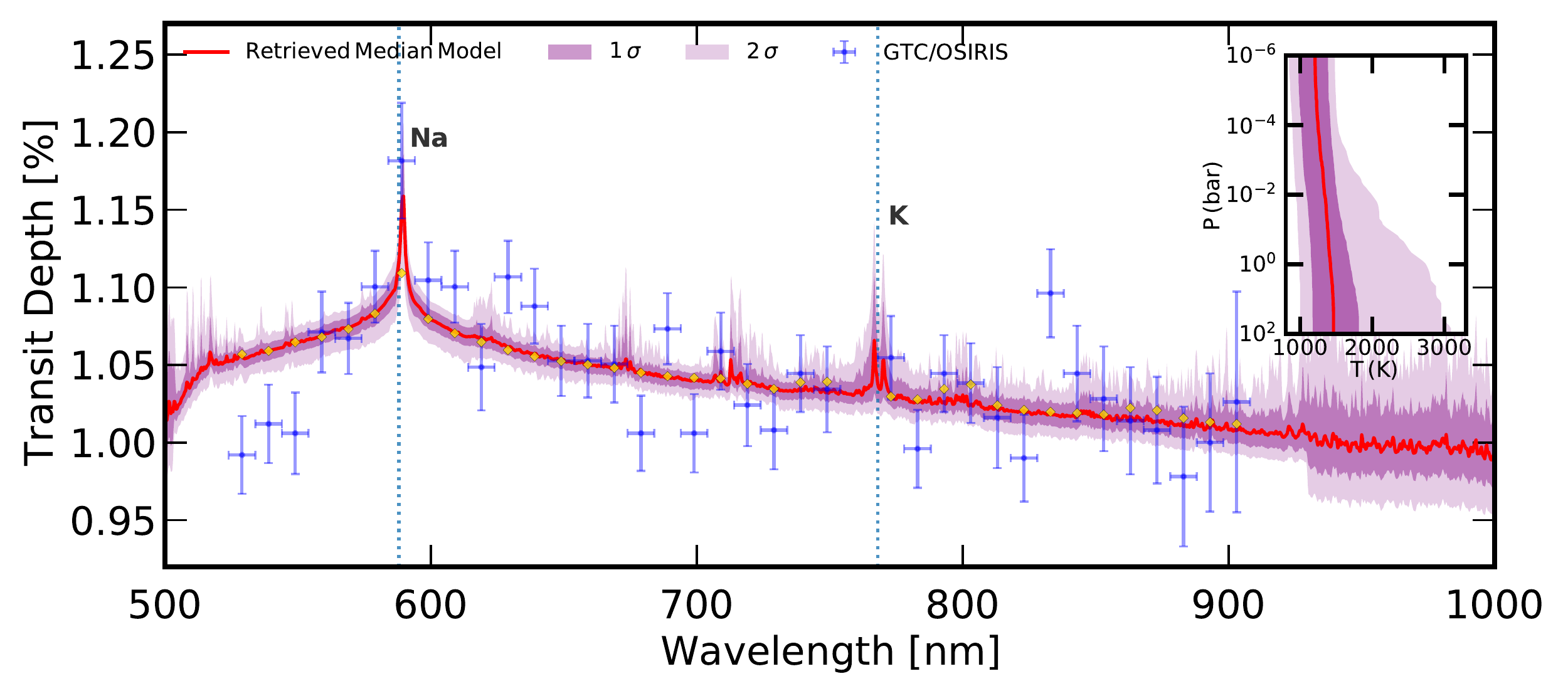}
\caption{\object{WASP-21b}'s transmission spectrum and retrieved models. The blue crosses are the error bars and bin width of the observed spectrum by GTC/OSIRIS. It shows a strong absorption peak at $\sim$589~nm. This feature is preferentially explained by our retrieval models by including the opacity resulting from Na. We show in an inset the retrieved P-T profile for the data using our fiducial model. The red curve shows the retrieved median model while the purple shaded areas show the 1$\sigma$ and 2$\sigma$ confidence regions. The yellow diamonds show the binned version for the retrieved median model.} 
\label{fig:spectrum}
\end{figure*}

The transit parameters derived from the white-color light curve are given in Table~\ref{tab:gtc_param}. The derived GTC/OSIRIS transmission spectrum is presented in Table~\ref{tab:gtc_transpec}. The white-color light curve and spectral light curves are shown in Fig.~\ref{fig:GTCWhiteLC} and Fig.~\ref{fig:GTCSpecLC}, respectively. The resulting standard deviation of the best-fitting light-curve residuals is 532~ppm, which is 6.1 times the photon noise. In contrast, the standard deviation of the spectral light-curve residuals is 1.07--1.44 times the photon noise. The large value above photon noise for the white-color light curve is likely a result of the poor weather conditions (e.g., cloud crossing as seen in the raw flux time series). The time-dependent light-curve scatter is mostly common-mode, which was artificially reduced in the corrected spectral light curves when removing the common-mode trend.

\subsection{GTC/OSIRIS transmission spectrum}
\subsubsection{Spectral retrieval analysis}
\label{sec:retrieval}

\begin{table*}
  \centering
  \caption{Summary of GTC/OSIRIS retrieval results. }
  \label{tab:retrievals}
  \begin{tabular}{lccccc}
  \hline\hline\noalign{\smallskip}
  Model     &$\log_{10}(X_{\text{Na}})$ & Na detection significance &$\log_{10}(X_{\text{K}})$& $\ln(\mathcal{Z})$& $\bar{\chi}^2$\\ \noalign{\smallskip}
  \hline\noalign{\smallskip}
  Fiducial     & $-2.57 ^{+ 0.84 }_{- 1.24 }$ & 3.5-$\sigma$& $-7.42 ^{+ 2.22 }_{- 2.65 }$ &235.79 & 2.25 \\ \noalign{\smallskip}
  Simplified   & $-3.31 ^{+ 1.34 }_{- 1.77 }$ & 4.9-$\sigma$ & $-7.53 ^{+ 2.26 }_{- 2.80 }$ & 239.12 & 1.42\\ \noalign{\smallskip}
  \hline\noalign{\smallskip}
  \end{tabular}
\end{table*}

The GTC transmission spectrum of WASP--21b is shown in Figure \ref{fig:spectrum}. We performed a retrieval analysis to constrain the atmospheric properties of the planet at the day-night terminator region. Our atmospheric retrieval code was adapted from the works of \citet{Pinhas2018} as used in previous studies \citep[e.g.,][]{2018A&A...616A.145C, 2019ApJ...887L..20W}. Our code computes line-by-line radiative transfer in a transmission geometry assuming a plane parallel planetary atmosphere in hydrostatic equilibrium. Our model retrieves the pressure-temperature (P-T) profile of the atmosphere utilizing the six-parameter prescription of \citet{2009ApJ...707...24M} in an atmosphere that spans from $10^2$ to $10^{-6}$ bar. 

In the retrieval framework the volumetric mixing ratios of the chemical species in the atmosphere are free parameters and assumed to be constant. We consider absorption due to molecules and atomic species that could be present in hot Jupiter atmospheres \citep{2016SSRv..205..285M}. The chemical opacity sources considered in this work are H$_2$-H$_2$ and H$_2$-He collision induced absorption \citep[CIA;][]{Richard2012}, CH$_4$ \citep{Yurchenko2014}, CO \citep{Rothman2010}, CO$_2$ \citep{Rothman2010}, H$_2$O \citep{Rothman2010}, HCN \citep{Barber2014}, K \citep{Allard2016}, Na \citep{Allard2019}, NH$_3$ \citep{Yurchenko2011}, TiO \citep{Schwenke1998}, AlO \citep{Patrascu2015}, and VO \citep{McKemmish2016}. The opacities for the chemical species are computed following the methods of \citet{2017MNRAS.472.2334G} and with H$_2$-broadened Na and K cross sections as explained in \citet{2019ApJ...887L..20W}.

Our models consider the possibility of inhomogeneous cloud and haze cover using the parametrization of \citet{2017MNRAS.469.1979M}. The model considers cloudy regions of the atmosphere to consist of an opaque cloud deck with a cloud-top pressure $P_\mathrm{cloud}$ in units of bar and scattering due to hazes above the clouds. In the parametrization, hazes are included as $\sigma=a\sigma_0(\lambda/\lambda_0)^\gamma$, where $\gamma$ is the scattering slope, $a$ is the Rayleigh-enhancement factor, and $\sigma_0$ is the H$_2$ Rayleigh scattering cross-section at a reference wavelength. The inhomogeneous clouds and scattering hazes are included through the parameter $\bar{\phi}$, which is the cloud/haze fraction cover in the planet's atmosphere. The Bayesian inference and parameter estimation is conducted using the nested sampling algorithm implemented via the MultiNest application \citep{2009MNRAS.398.1601F} through the Python interface PyMultiNest \citep{2014A&A...564A.125B}. 

We perform an initial exploratory retrieval considering possible absorption due to all 11 chemical species considered in this work, inhomogeneous cloud and haze cover, and a parametric P-T profile. As expected, we find that the only chemical species relatively constrained by the data are those with strong spectroscopic signatures in the optical wavelengths. Therefore we determine our fiducial model to consider absorption due to H$_2$O, Na, K, TiO, AlO and VO only, as well as a parametric P-T profile and inhomogeneous cloud and haze cover. The fiducial model has a total of 17 parameters: 6 chemical abundances, 6 parameters for the P-T profile, 4 parameters for the cloud and haze prescription, and 1 parameter for the reference pressure corresponding to the reference planetary radius of 1.162 $R_J$.

\begin{figure}
\centering
\includegraphics[width=1\linewidth]{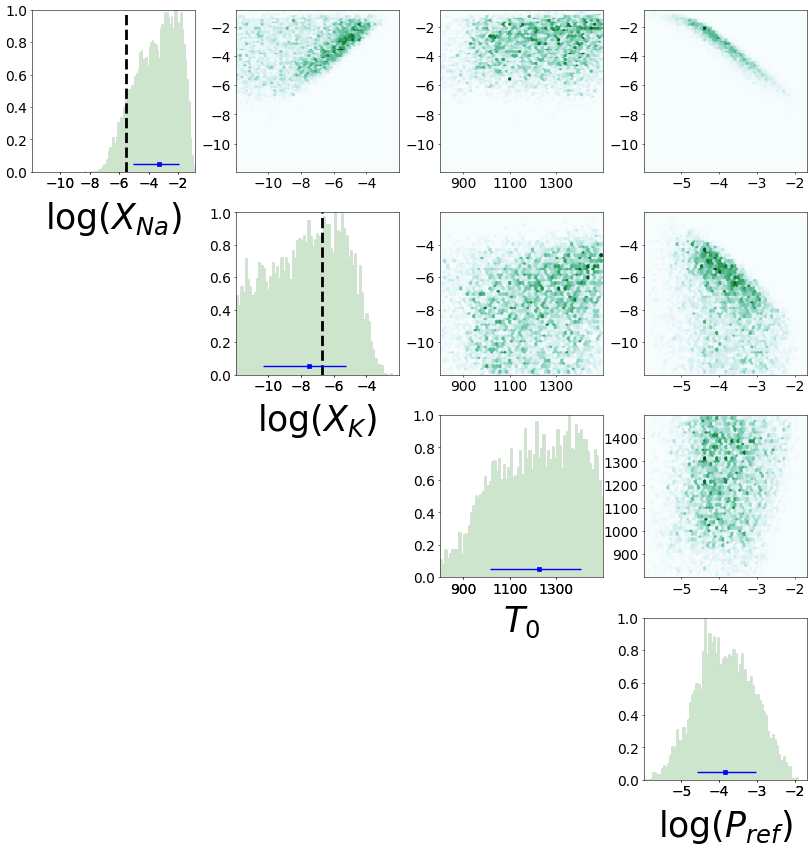}
\caption{Posterior distribution for the simplified model retrieval of \object{WASP-21b}'s transmission spectrum. 
This four parameter model is the highest evidence model and considers absorption due to Na and K only in a clear isothermal atmosphere. Na and K solar abundance expectations are shown using black dashed lines.} 
\label{fig:posterior_distribution}
\end{figure}

Figure \ref{fig:spectrum} shows the retrieved median model to the observations along with the 1$\sigma$ and 2$\sigma$ confidence regions. It also shows in an inset the retrieved pressure-temperature profile for the fiducial model. Our results suggest that the features in the spectrum can be explained by the presence of Na in the planet's atmosphere. Using the fiducial model as reference we report a possible detection of Na at a confidence level of 3.5-$\sigma$. The retrieved Na abundance is $\log_{10}(X_{\text{Na}})=-2.57 ^{+ 0.84 }_{- 1.24 }$. Besides Na, K also shows possible spectroscopic signatures at $\sim$770 nm. Our models do not exhibit a strong preference for the presence of K in the spectrum of WASP-21b and derive a largely unconstrained abundance of $\log_{10}(X_{\text{K}})=-7.42 ^{+ 2.22 }_{- 2.65 }$. Similarly, the data does not provide strong constraints on the P-T profile or the presence of clouds and hazes in the atmosphere of WASP-21b. The retrieved cloud and haze parameters are unconstrained and consistent with a mostly clear atmosphere, partly due to the lack of features in the data indicating a scattering slope. The P-T profile remains largely unconstrained with a derived temperature at 100~mbar, close to the photosphere, of $T_\mathrm{100~mbar}=1371^{+254}_{-230}$ K consistent with the equilibrium temperature of the planet.

Given that current spectroscopic observations do not place strong constraints on the P-T profile or the presence of clouds and hazes we consider a simplified model retrieval. The simplified model considers absorption due to Na and K only, an isothermal P-T profile, and a clear atmosphere. The simplified model retrieved abundances are $\log_{10}(X_{\text{Na}})=-3.31 ^{+ 1.34 }_{- 1.77 }$ and $\log_{10}(X_{\text{K}})=-7.53 ^{+ 2.26 }_{- 2.80 }$, consistent with the fiducial model. The derived Na abundance is marginally consistent with expectations from solar abundance chemistry of $\log_{10}(\text{Na}/\text{H}) = -5.76$ \citep{Asplund2009}. Similarly, the retrieved temperature for the isotherm is $T=1224 ^{+ 181 }_{- 208 }$~K, consistent with the equilibrium temperature and the derived temperature at 100~mbar in the fiducial model. Using this simplified model as reference, Na is detected at a confidence level of 4.9-$\sigma$. The posterior distribution for this retrieval is shown in Fig. \ref{fig:posterior_distribution} and our retrieval results summarized in Table \ref{tab:retrievals}.

\subsubsection{A search of narrow alkali features}

We also examined the 16~$\AA$ bin transmission spectrum zoomed at the Na and K doublets (see Fig.~\ref{fig:GTC_NaK} in Appendix \ref{sec:app_lowres}). The narrow-band transmission spectrum does hint signs of excess absorption at the cores of the Na doublet and the K D$_1$ line, but not at significant levels. The K D$_1$ line is $\Delta R_\mathrm{p}/R_\star=0.0069\pm 0.0035$ higher than the weighted mean of the two neighboring bands. The K D$_2$ line is unfortunately located in the telluric oxygen-A band. High-resolution transmission spectroscopy from ultra-stable radial velocity spectrographs like ESPRESSO is required to confirm this tentative evidence of excess absorption.

\section{High-resolution data analysis}
\label{sec:hires_analysis}

\subsection{Rossiter-McLaughlin effect in radial velocities}
\label{sec:rm_analysis}

\begin{figure}
\centering
\includegraphics[width=1.0\linewidth]{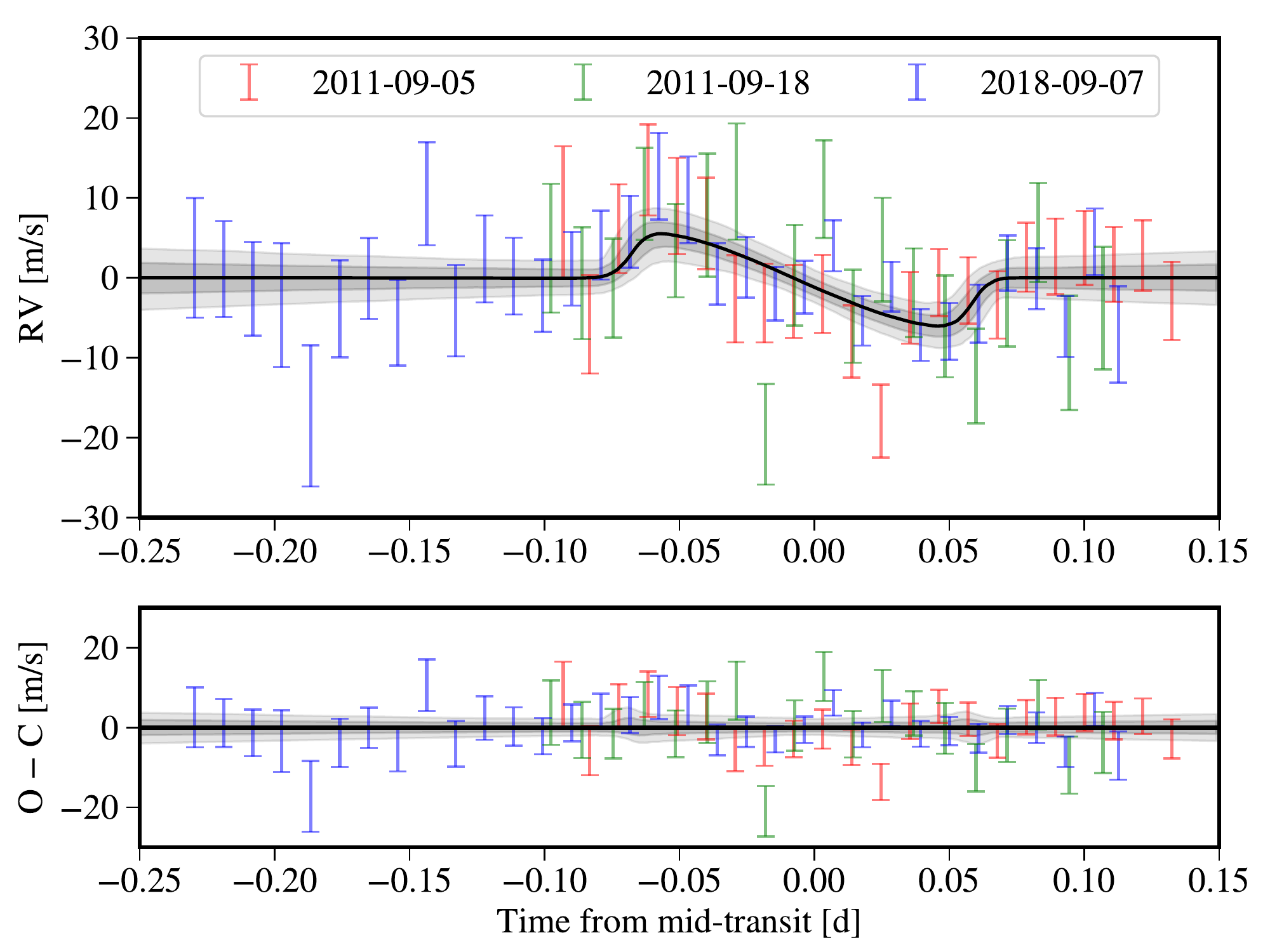}
\caption{The Rossiter-McLaughlin effect in the radial velocity curve of WASP-21. The transits observed on the nights of September 5, 2011, September 18, 2011, and September 7, 2018 are shown in red, green, and blue, respectively. The median RM model and its 1-$\sigma$ and 2-$\sigma$ confidence regions are shown in black and gray. Top panel shows the anomaly caused by the RM effect. Bottom panel shows the residuals after removing the median RM model.} 
\label{fig:rm}
\end{figure}

\begin{table}
     \centering
     \caption{Derived parameters from Rossiter-McLaughlin effect.}
     \label{tab:rm_param}
     \begin{tabular}{lr}
     \hline\hline\noalign{\smallskip}
     Parameter & Value \\\noalign{\smallskip}
     \hline\noalign{\smallskip}
        Systemic velocity \#1, $\gamma_1$ [$\mathrm{km\,s}^{-1}$]         & $-89.4499 \pm 0.0013$ \\
        Systemic velocity \#2, $\gamma_2$ [$\mathrm{km\,s}^{-1}$]         & $-89.4484 \pm 0.0020$ \\
        Systemic velocity \#3, $\gamma_3$ [$\mathrm{km\,s}^{-1}$]         & $-89.4365 \pm 0.0010$ \\
        Proj. rotation velocity, $v\sin i_\star$ [$\mathrm{km\,s}^{-1}$]  & $0.66 \pm 0.14$     \\
        Proj. spin-orbit angle, $\lambda$ [deg]                           & $8 ^{+26}_{-27}$\\
        RV semi-amplitude, $K_\star$ [$\mathrm{km\,s}^{-1}$]              & $0.0325 \pm 0.0053$  \\
        Offset to expected mid-transit, $\Delta T_\mathrm{C}$ [d]         & $-0.0047 \pm 0.0042$ \\
        Limb darkening coeff., $u_1$, $u_2$                               & 0.3991, 0.2830    \\
    \hline\noalign{\smallskip}
    \end{tabular}
\end{table}

The transit of a planet, blocking light from a part of the stellar disk, would introduce an asymmetric distortion in the line profiles of the stellar spectrum. The resulting radial velocity (RV) curves will exhibit an apparent anomaly known as the Rossiter-McLaughlin \citep[RM;][]{1924ApJ....60...15R,1924ApJ....60...22M} effect. We collected the RVs from the three transits measured by the HARPS and HARPS-N pipelines, and jointly fit them with the ARoME library \citep{2013A&A...550A..53B} implemented by a Python interface\footnote{\url{https://github.com/andres-jordan/PyARoME}}. ARoME can appropriately model the RM effect for the RVs derived from the CCF-based approach (e.g., HARPS). A circular orbit was adopted for WASP-21b \citep{2010A&A...519A..98B}. The combined RV model can be written as: 
\begin{equation}
v_\star=v_\mathrm{RM}+K_\star\sin [2\pi(t-T_\mathrm{C})/P]+\gamma, 
\end{equation}
where $K_\star$ is the stellar RV semi-amplitude, $T_\mathrm{C}$ is the mid-transit time, $P$ is the orbital period, and $\gamma$ is the systemic RV. The RM anomaly $v_\mathrm{RM}$ is described by ARoME, which contains the following parameters: orbital period $P$, mid-transit time $T_\mathrm{C}$, scaled semi-major axis $a/R_\star$, orbital inclination $i$, planet-to-star radius ratio $R_\mathrm{p}/R_\star$, projected stellar rotation velocity $v\sin i_\star$, projected spin-orbit angle $\lambda$, and quadratic limb-darkening coefficients ($u_1$, $u_2$). ARoME requires additional three parameters to define the line profile of CCF, that is, the width of a non-rotating star $\beta_0$ (adopted as 1.3~km\,s$^{-1}$), the width of the best Gaussian fit to out-of-transit CCF $\sigma_0$ (adopted as 2.9~km\,s$^{-1}$), and stellar macro-turbulence $\zeta$ (adopted as 2.3~km\,s$^{-1}$)

In the joint fitting of RV curves, the three transits shared the same values for $K_\star$, $\Delta T_C$, $v\sin i_\star$, and $\lambda$, but were allowed to have different values for $\gamma$. The offset to the predicted mid-transit time, $\Delta T_C$, was calculated based on the ephemeris listed in Table \ref{tab:gtc_param}, assuming that there is no transit timing variation. A Gaussian prior was imposed on $\Delta T_C$, which has a width of three times the error propagation from the ephemeris. The values of $P$, $a/R_\star$, $i$, and $R_\mathrm{p}/R_\star$ were fixed to those listed in Table \ref{tab:gtc_param}, and the limb-darkening coefficients were fixed to theoretical values (see Sect.~\ref{sec:gtclc}). We employed \texttt{emcee} \citep{2013PASP..125..306F} to perform the MCMC process to search for the best-fit solutions. A rescaling multiple factor $f_s$ for each night was used in the likelihood function to account for the over- or underestimation of error bars. In the end, the derived rescaling multiple factor was 1.07 $^{+0.20}_{-0.16}$ (transit \#1), 1.25 $^{+0.25}_{-0.19}$ (transit \#2), and 0.96 $^{+0.14}_{-0.11}$ (transit \#3), respectively.

The joint analysis of the three transits resulted in a projected spin-orbit angle of $8^\circ\,^{+26}_{-27}$. Stars with photospheres cooler than $\sim$6100~K in general have low obliquities \citep{2015ARA&A..53..409W}. WASP-21 has an effective temperature of $T_\mathrm{eff}=5800\pm 100$~K \citep{2010A&A...519A..98B}. The currently measured projected spin-orbit angle of $8^\circ\,^{+26}_{-27}$ seems to make WASP-21 follow this trend, but cautions should be taken given the large uncertainties. The derived stellar RV semi-amplitude of $32.5\pm 5.3$~m\,s$^{-1}$ is consistent with that derived from full-phase coverage RV curve without the RM anomaly in the discovery paper \citep[$37.2\pm 1.1$~$\mathrm{m\,s^{-1}}$;][]{2010A&A...519A..98B}. The projected stellar rotation velocity measured from the RM effect ($0.7\pm 0.1$~km\,s$^{-1}$) is smaller than the spectroscopically derived value \citep[$1.5\pm 0.6$~$\mathrm{km\,s^{-1}}$;][]{2010A&A...519A..98B}. The discrepancy in the values of $v\sin i_\star$ based on different methods have also been noticed in other planets and discussed by \citet{2017MNRAS.464..810B} and \citet{2018A&A...619A.150O}. The derived parameters are presented in Table \ref{tab:rm_param}. As shown in Fig.~\ref{fig:rm}, the RV anomaly caused by the RM effect is relatively small.

\subsection{Transmission spectroscopy of atomic lines}
\label{sec:line_ts}

\begin{figure}
\centering
\includegraphics[width=0.8\linewidth]{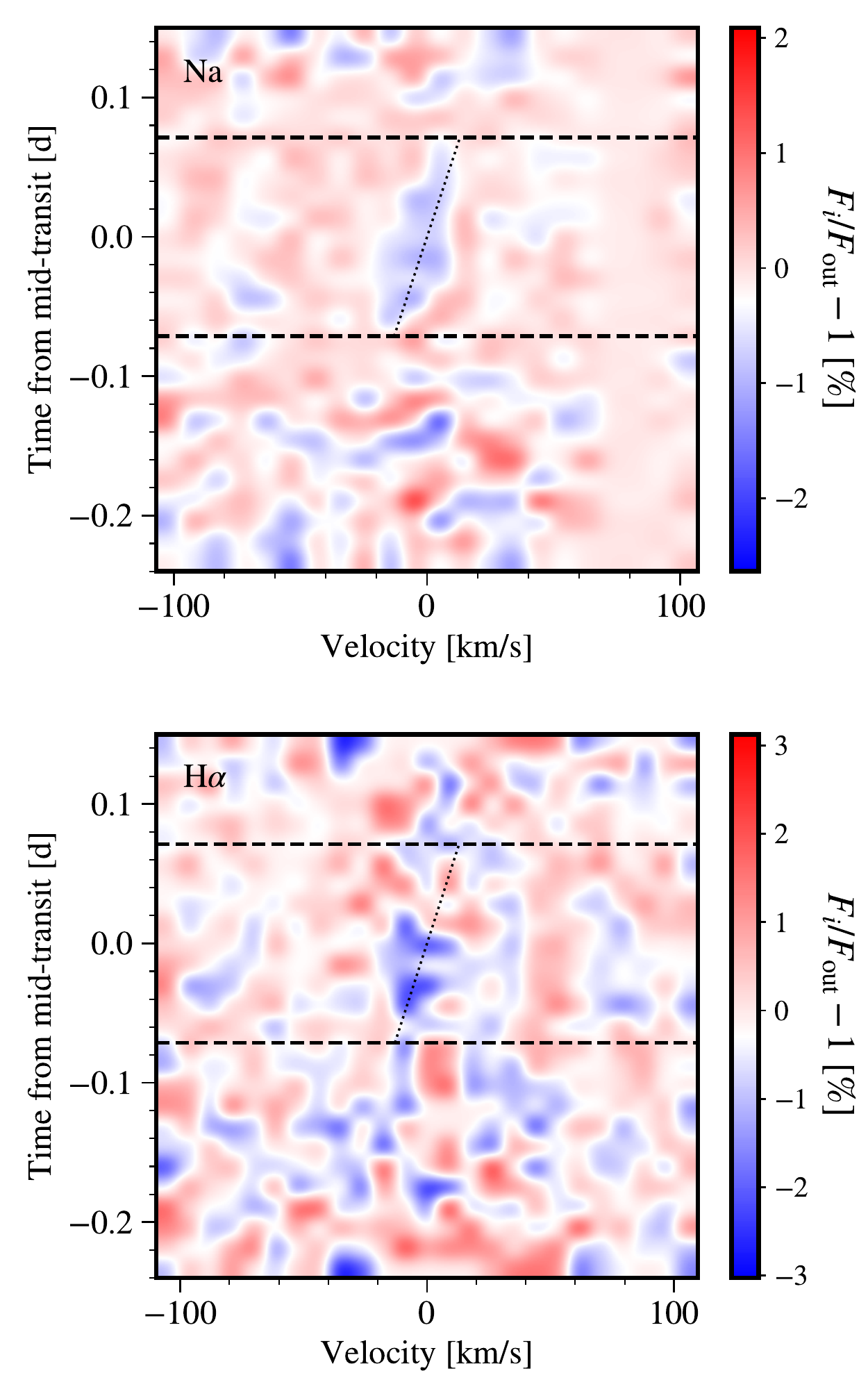}
\caption{Phase-resolved transmission spectrum at the Na ({\it top}) and H$\alpha$ ({\it bottom}) lines for WASP-21b. A radial velocity shift traced by the Na line center can be noticed during the transit event, which agrees with the radial velocity shift induced by the planet orbital motion. It is not clear at the H$\alpha$ line center.} 
\label{fig:transpec_2dmap}
\end{figure}

\begin{figure}
\centering
\includegraphics[width=0.8\linewidth]{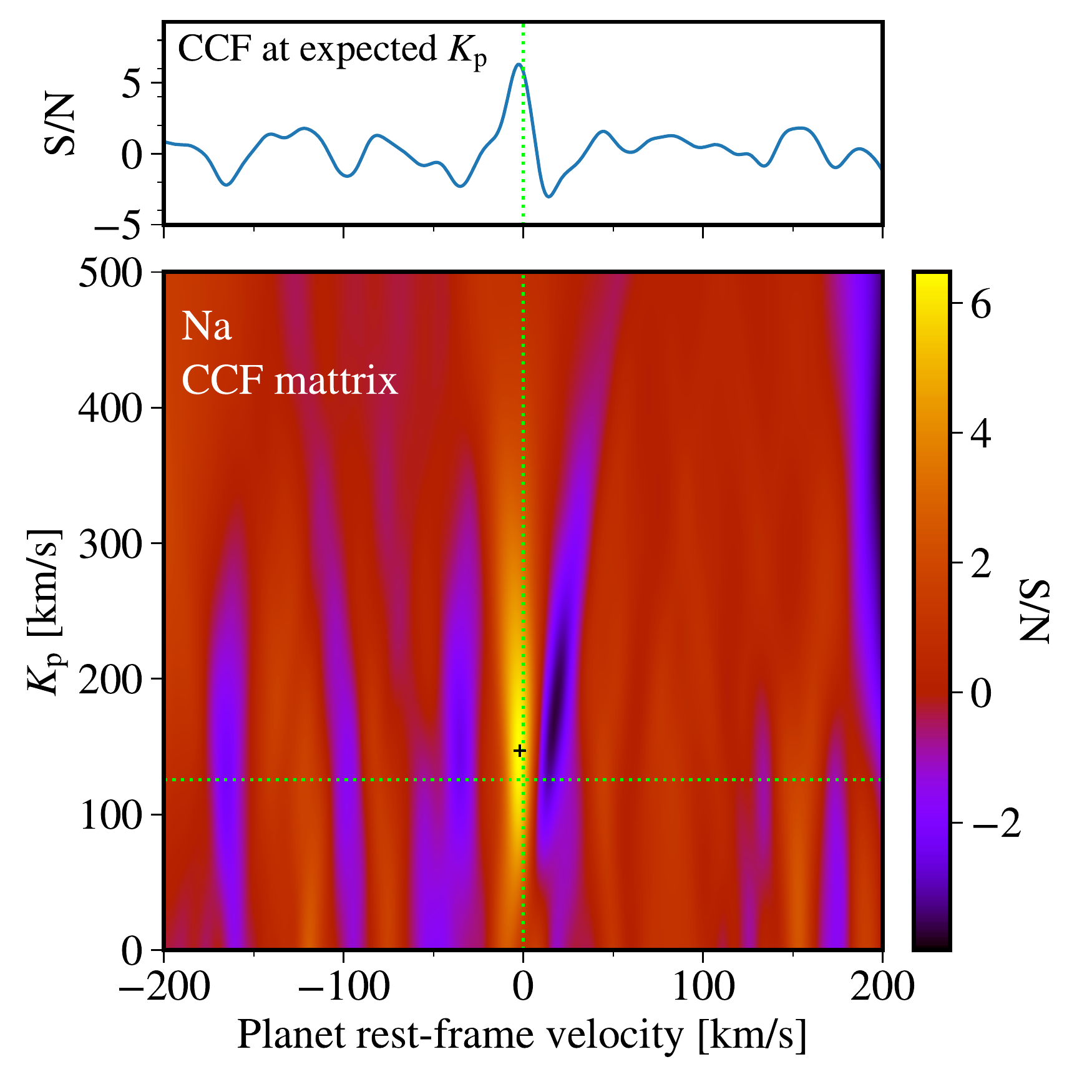}
\caption{Combined cross-correlation function (CCF) as a function of planet radial velocity (RV) semi-amplitude and RV shift of the Na doublet. The CCF has been expressed in the form of signal-to-noise ratio (S/N). The green dotted lines indicate the expected RV semi-amplitude and zero velocity. The black plus sign shows the location of the maximum CCF. {\it Top panel} shows the CCF at the expected RV semi-amplitude, corresponding to the horizontal green dotted line.} 
\label{fig:ccmap_Na}
\end{figure}

We followed the methodology detailed in \citet{2020A&A...635A.171C} to remove telluric and stellar lines in the acquired HARPS-N and HARPS spectra. In brief, the HARPS-N and HARPS spectra were first shifted back to the Earth's rest frame. Telluric H$_2$O and O$_2$ absorption lines were modeled and removed using the ESO software \texttt{molecfit} version 1.5.7 \citep{2015A&A...576A..77S,2015A&A...576A..78K}. The telluric corrected spectra were then shifted to the stellar rest frame using barycentric Earth radial velocities and stellar radial velocities without the RM anomaly. The spectra were normalized and divided by the out-of-transit master (hereafter master-out) spectrum to remove stellar lines on a nightly basis. In this process, the continuum variation between individual exposures was corrected using a fourth-order polynomial function fitted on the individual-to-master-out flux-ratio spectra. And the master-out was constructed as the weighted mean of the out-of-transit spectra, using the square of S/N as weights. After subtracting a value of one, the resulting residual spectra matrix is equivalent to a phase-resolved transmission spectrum. 

The residual spectra in principle still contain distortion features of stellar lines introduced by the transit of a planet, during which a part of emergent stellar flux or radial velocity component are blocked. The \texttt{Spectroscopy Made Easy} tool \citep[SME;][]{1996A&AS..118..595V,2017A&A...597A..16P} was employed to create the models for center-to-limb variation (CLV) and Rossiter-McLaughlin (RM) effect. The simulated line distortion features caused by the CLV and RM effect were then corrected in the residual spectra. In the case of WASP-21, the CLV and RM effects are small and negligible (see Fig.~\ref{fig:NaHa_CLV_RM} and Fig.~\ref{fig:activity_indicators}). However, for completeness, the subsequent analysis has included the correction of both effects.

Figure \ref{fig:transpec_2dmap} presents the phase-resolved transmission spectrum at the Na and H$\alpha$ lines in the stellar rest frame. To enhance S/N, the residual spectra of three nights have been combined and also binned in both time domain and velocity domain. Although the S/N is limited, it is still noticeable that the Na line center exhibits an excess absorption during the transit. The RV shift of the excess Na absorption during the transit is consistent with that induced by the planet orbital motion, indicative of a planetary origin. In contrast, the H$\alpha$ line center is dominated by noise. 

In order to quantitatively assess the RV shift of the Na doublet excess absorption, the ideal way is to simultaneously fit the absorption profile at each time grid \citep[e.g.,][]{2018NatAs...2..714Y,2019A&A...628A...9C}. Given the low S/N of our current data set, we chose to perform the cross-correlation technique on the unbinned residual spectra using the best-fit doublet absorption model obtained on the combined transmission spectrum. The model template was shifted by a value of $\Delta v_\mathrm{abs}$ before its cross-correlation with the residual spectra that were fully in-transit. The cross-correlation functions (CCF) were then combined after they were shifted to planet rest frame using a grid of planet RV semi-amplitude values ($K_\mathrm{p}$). The combined CCF at grids of $K_\mathrm{p}$ and $\Delta v_\mathrm{abs}$ is shown in Fig.~\ref{fig:ccmap_Na}, which is expressed in the form of S/N, that is, the CCF map has been normalized by the standard deviation of $|\Delta v_\mathrm{abs}|=100-200$~km\,s$^{-1}$ at a given $K_\mathrm{p}$. The CCF map shows a maximum S/N of 6.5 at the location of $K_\mathrm{p}=147^{+61}_{-52}$~km\,s$^{-1}$ and $\Delta v_\mathrm{abs}=-2^{+3}_{-4}$~km\,s$^{-1}$. The value of $K_\mathrm{p}$ is consistent with the predicted value of $K_\mathrm{p}=2\pi a\sin(i)/P=125.6\pm3.8$~km\,s$^{-1}$ (assuming zero eccentricity and adopting parameters in Table \ref{tab:adopted_param}), revealing that this excess absorption signal is related to the planet.

\begin{figure*}
\centering
\includegraphics[width=1.0\linewidth]{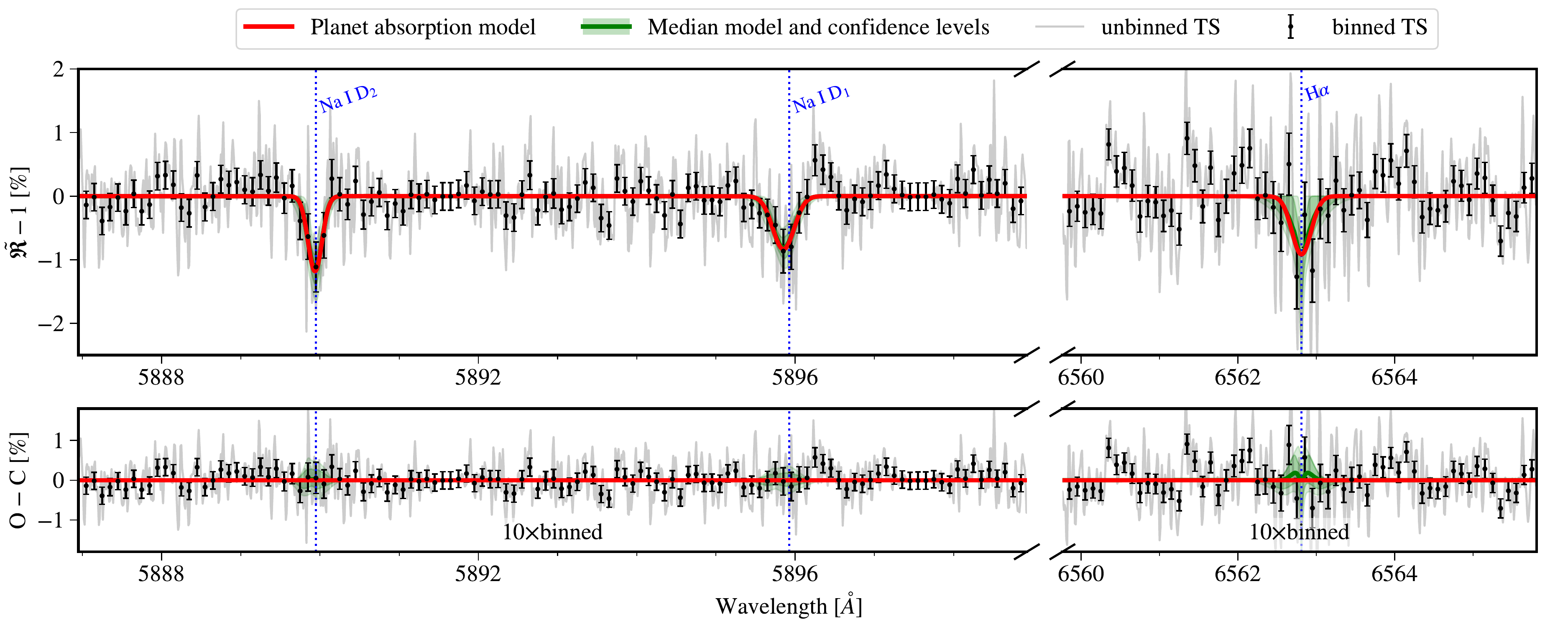}
\caption{High-resolution transmission spectrum of WASP-21b at the Na doublet and H$\alpha$ lines. The black circles with error bars show the 10$\times$binned transmission spectrum, while the gray line shows the 0.01~$\AA$ version. The red line presents the best-fit Gaussian function to the Na doublet and H$\alpha$ lines. The green lines and shaded areas refer to the median and 1$\sigma$ confidence region of the Gaussian fit. {\it Top panel} shows the transmission spectrum, while {\it bottom panel} shows the one after subtracting the best-fit model.} 
\label{fig:hires_transpec}
\end{figure*}

We created the final transmission spectrum at the atomic lines of interest (e.g., Na, H$\alpha$, etc) by weighted combining the residual spectra obtained during fully in-transit, which have been shifted to the planet rest frame before the combination. The RV shift was performed using the expected planet RV semi-amplitude ($K_\mathrm{p}=125.6$~km\,s$^{-1}$). We fitted a Gaussian function to each atomic line to retrieve the line parameters such as contrast, center, and FWHM. The fitting procedure was implemented by the \texttt{emcee} package \citep{2013PASP..125..306F} with an error rescaling multiple, and the results were given in Table \ref{tab:line_param}. The transmission spectrum at the Na doublet and H$\alpha$ lines is presented in Fig.~\ref{fig:hires_transpec}, along with the best-fit models. 

\begin{table}
  \centering
  \caption{Parameters from the Gaussian fit to the Na line profile. }
  \label{tab:line_param}
  \begin{tabular}{lccc}
  \hline\hline\noalign{\smallskip}
  Parameter     & Unit & Na D$_1$ & Na D$_2$\\ \noalign{\smallskip}
  \hline\noalign{\smallskip}
  \multicolumn{4}{c}{\it Free parameters}\\\noalign{\smallskip}
  Line contrast & \%     & $0.84^{+0.16}_{-0.17}$ & $1.18^{+0.23}_{-0.24}$ \\ \noalign{\smallskip}
  Line center   & $\AA$  & $5895.851^{+0.029}_{-0.032}$ & $5889.938^{+0.020}_{-0.021}$ \\ \noalign{\smallskip}
  Line FWHM     & $\AA$  & $0.317^{+0.056}_{-0.050}$ & $0.207^{+0.043}_{-0.039}$ \\ \noalign{\smallskip}
  Error multiple   & --  & \multicolumn{2}{c}{$0.616^{+0.013}_{-0.013}$} \\ \noalign{\smallskip}
  \multicolumn{4}{c}{\it Derived parameters}\\\noalign{\smallskip}
  Effective radius & $R_\mathrm{p}$     & $1.307^{+0.054}_{-0.053}$ & $1.414^{+0.072}_{-0.070}$ \\ \noalign{\smallskip}
  Center offset    & km\,s$^{-1}$       & $-3.7^{+1.5}_{-1.6}$ & $-0.7^{+1.0}_{-1.1}$ \\ \noalign{\smallskip}
  Line FWHM        & km\,s$^{-1}$       & $16.1^{+2.9}_{-2.5}$ & $10.5^{+2.2}_{-2.0}$ \\ \noalign{\smallskip}
  \hline\noalign{\smallskip}
  \end{tabular}
\end{table}

Excess absorption was only detected at the Na doublet. The derived line contrast is $0.84^{+0.16}_{-0.17}$\% for Na D$_1$ and $1.18^{+0.23}_{-0.24}$\% for Na D$_2$, respectively, resulting in a line ratio of $h_\mathrm{D_2}/h_\mathrm{D_1}=1.41^{+0.46}_{-0.34}$, which is in line with previous studies on the other hot Jupiters \citep[e.g.,][except for WASP-76b]{2019AJ....158..120Z}. The effective radius at the line center can be converted from the line contrast $h$ as $R_\mathrm{eff}=[1+h/\mathcal{D}]^{1/2}$, where we adopted for $\mathcal{D}=R_\mathrm{p}^2/R_\star^2$ the value of the 10~nm GTC/OSIRIS bandpass that Na was located in ($R_\mathrm{p}/R_\star=0.1087 ^{+0.0017 }_{-0.0017 }$). The effective radius at the centers of Na D$_1$ ($1.307^{+0.054}_{-0.053}$~$R_\mathrm{p}$) and D$_2$ ($1.414^{+0.072}_{-0.070}$~$R_\mathrm{p}$) are well below the Roche lobe radius $3.05^{+0.17}_{-0.16}$~$R_\mathrm{p}$, calculated according to \citet{1983ApJ...268..368E}. The line center of the Na doublet shows an average net blueshift of $-1.6\pm 0.9$~km\,s$^{-1}$, which might indicate a day-to-night planetary wind.

We did not detect any significant excess absorption at the H$\alpha$ line. A Gaussian fit with the center fixed at the laboratory wavelength would result in an excess absorption of $<2.3\%$ at a confidence level of 95\%, which is a rather loose constraint and indicates a noisy transmission spectrum at the H$\alpha$ line. We have examined the transmission spectrum at some other atomic lines that could serve as stellar activity indicators \citep[e.g.,][]{2010MNRAS.403.2157H,2019MNRAS.490L..86Y}. None of them showed any significant excess absorption, nor showed any significant features introduced by CLV and RM (see e.g., Fig.~\ref{fig:activity_indicators}). Therefore, the observed excess absorption at the Na doublet is unlikely a result of stellar activity or the CLV and RM effects.

Since the final transmission spectrum was created using the fully in-transit residual spectra, we performed the empirical Monte Carlo simulations \citep[e.g.,][]{2008ApJ...673L..87R} to double check whether or not the excess absorption can only be detected in-transit. This was implemented by creating an in-transit (hereafter mock-in) and an out-of-transit mock data sets (hereafter mock-out), and performing the same analysis as the real data. Three scenarios were tested. In the ``in-in'' scenario, the real in-transit spectra were randomly divided into two groups, one as mock-in and the other as mock-out. In the ``out-out'' scenario, the real out-of-transit spectra were randomly divided. In the ``in-out'' scenario, mock-in spectra were randomly selected from the real in-transit spectra, while mock-out were from the real out-of-transit. The total number of mock-in and mock-out spectra was the same as that of the real data for both ``in-in'' and ``out-out'', while the number was randomly generated but kept being no less than half of the real number for ``in-out''. For all three scenarios, the mock in-to-out number ratio was kept being the same as the real number ratio. Once the two samples were ready, we created the mock transmission spectrum following the same way as aforementioned. We then measured the absorption depth within a defined passband bin centered at the target line on the transmission spectrum. 

\begin{figure}
\centering
\includegraphics[width=1.0\linewidth]{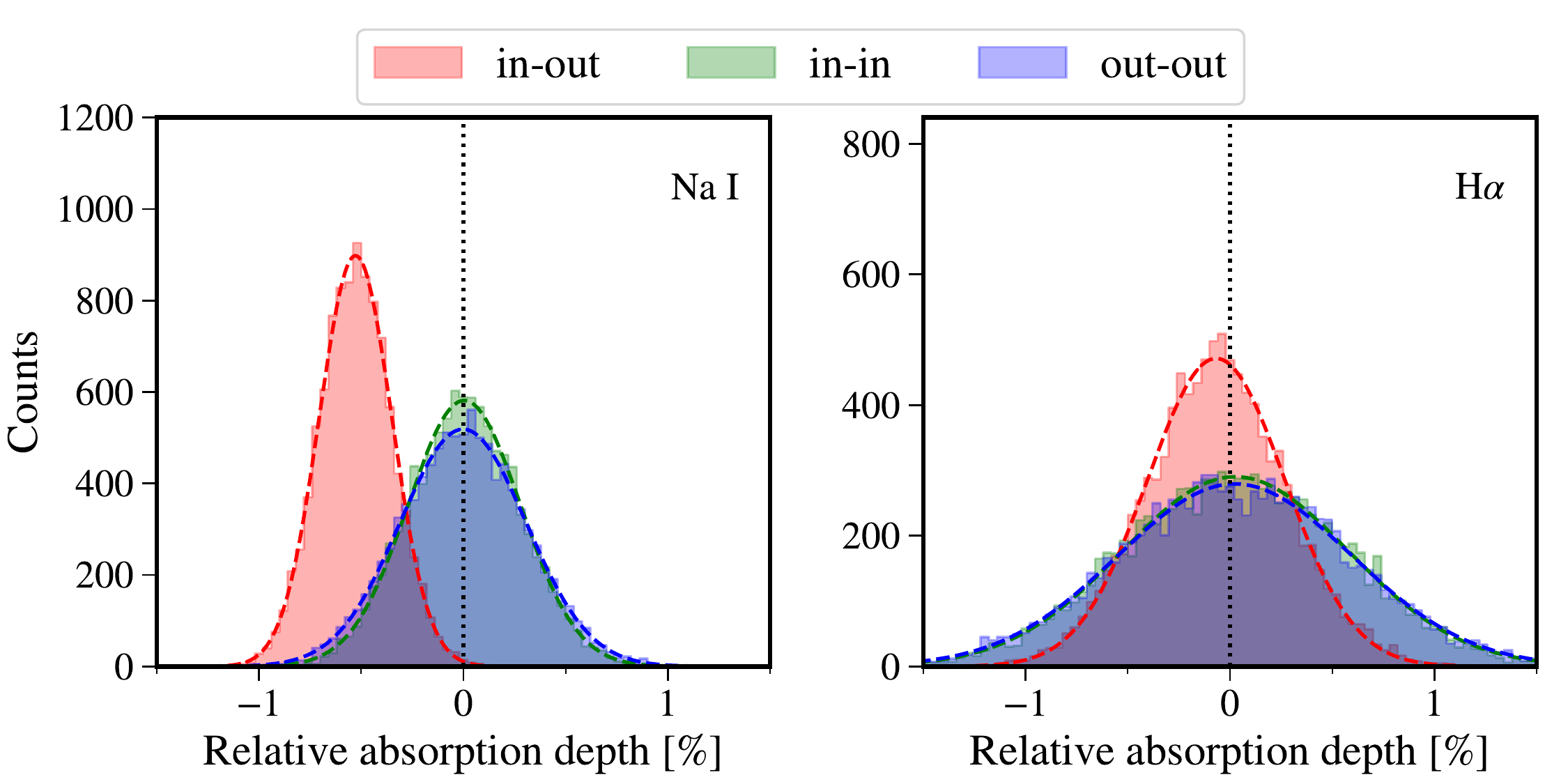}
\caption{Absorption depths measured in the mock transmission spectra created in the empirical Monte Carlo simulation ({\it left}: Na, {\it right}: H$\alpha$). The distribution of the scenarios ``in-out'', ``in-in'', and ``out-out'' are coded in red, green, blue colors, respectively. We refer the reader to Sect.~\ref{sec:line_ts} for details. } 
\label{fig:EMC}
\end{figure}

Figure \ref{fig:EMC} presents the resulting distributions for the Na doublet and H$\alpha$ lines. For the Na doublet, D$_1$ and D$_2$ were measured in two 0.35~$\AA$ passbands and averaged. We can only measure an excess absorption depth of $0.53\pm 0.21$\% in the ``in-out'' scenario, which is consistent with the value ($0.70\pm 0.13$\%) measured on the real data. The ``in-in'' and ``out-out'' scenarios have posterior distributions centered at zero, indicating that either the signals have been canceled out or there is no signal. This confirms that the detected excess Na signal can only be detected in-transit. For H$\alpha$, it was measured in a 0.75~$\AA$ band, consistent with zero.

\section{Discussion}
\label{sec:discuss}

\subsection{The atmosphere of WASP-21b}

With the flux calibrated low-resolution transmission spectrum acquired by GTC/OSIRIS, we have unambiguously detected a broad spectral signature centered at the Na doublet line. This signature indicates that the Na line wing is probably pressure-broadened, and that we are looking into relatively low altitudes of the atmosphere. The pressure broadening of the Na line has already been observed in several low-mass hot Jupiters, ranging from half-Jupiter-mass planets (WASP-96b, \citealt{2018Natur.557..526N}; XO-2b, \citealt{2019AJ....157...21P}) to Saturn-mass planets \citep[WASP-39b,][]{2016Natur.529...59S,2016ApJ...832..191N} and to sub-Saturn-mass planets \citep[WASP-127b,][]{2018A&A...616A.145C}. The transmission spectrum of WASP-127b also exhibits a pressure broadening at the K doublet that is stronger than Na \citep{2018A&A...616A.145C}. For WASP-21b, we are not able to detect any significant pressure broadening at the K doublet, but do see a tentative evidence of excess absorption at the K D$_1$ line. The inference of pressure broadening of alkali lines indicates that the atmosphere is at least partially clear, making the planet extremely favorable for further follow-up atmospheric characterization. 

The high-resolution transmission spectrum of WASP-21b further confirms the presence of Na in its atmosphere at higher altitudes. This is achieved by resolving the radial velocity shift of the excess absorption at the Na doublet line, which is only detectable during the transit and consistent with the planet orbital motion. The excess absorption extends $\sim$28$H$ at the Na D$_1$ line and $\sim$37$H$ at the Na D$_2$ line, where $H=k_\mathrm{B}T_\mathrm{eq}/(\mu g_\mathrm{p})=944$~km is the atmospheric scale height. This is much wider than the range covered by our GTC/OSIRIS low-resolution spectrum ($\sim$8$H$). At high resolution, the extension of the excess absorption at Na (doublet averaged) has been measured to vary between $\sim$16$H$ \citep[WASP-127b,][]{2019AJ....158..120Z} and $\sim$66$H$ \citep[WASP-49b,][]{2017A&A...602A..36W} for different hot Jupiters. 

The Na doublet of WASP-21b exhibits a tentative net blueshift of $-1.6\pm 0.9$~km\,s$^{-1}$, indicative of a possible day-to-night planetary wind. This value is similar to wind velocities measured in the hot Jupiters HD 189733b \citep[$-1.9^{+0.7}_{-0.6}$~km\,s$^{-1}$,][]{2015ApJ...814L..24L} and WASP-49b \citep[$-1.7\pm 1.6$~km\,s$^{-1}$,][]{2017A&A...602A..36W}, which are also traced by Na. The Na-traced net velocities have also been reported for the hot Jupiter WASP-52b \citep[$-0.6\pm 0.7$~km\,s$^{-1}$,][]{2020A&A...635A.171C}, and the ultra hot Jupiters KELT-9b \citep[$+2.9\pm 1.0$~km\,s$^{-1}$,][]{2019A&A...627A.165H} and MASCARA-2b \citep[$-3.1\pm 0.4$~km\,s$^{-1}$,][]{2019A&A...628A...9C}. All of them refer to the doublet averaged value. The wind velocity and its variability, when measured precisely and phase-resolved, could be related to global circulation and drag strength \citep[e.g.,][]{2013ApJ...762...24S,2014ApJ...795...24K,2019AJ....157..209F,2020ApJ...888....2K}.

\subsection{Pressure broadening in the atmospheres of hot Jupiters}

\begin{figure*}
\centering
\includegraphics[width=1\linewidth]{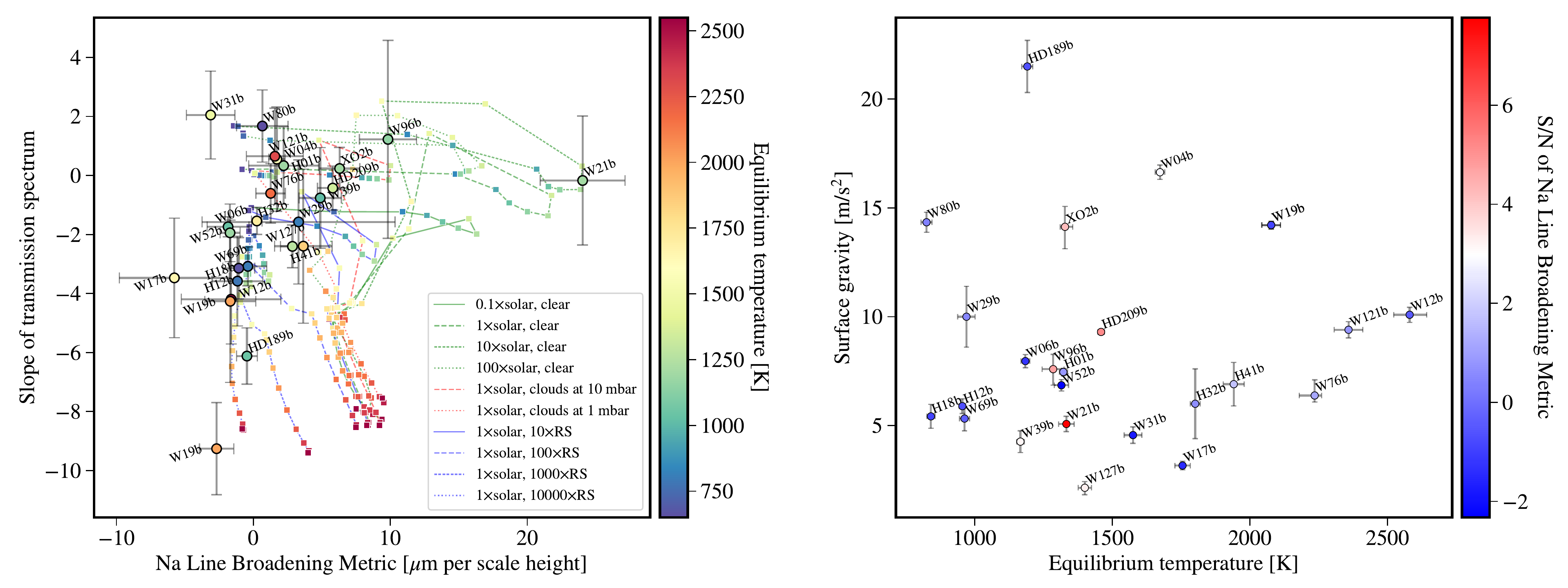}
\includegraphics[width=1\linewidth]{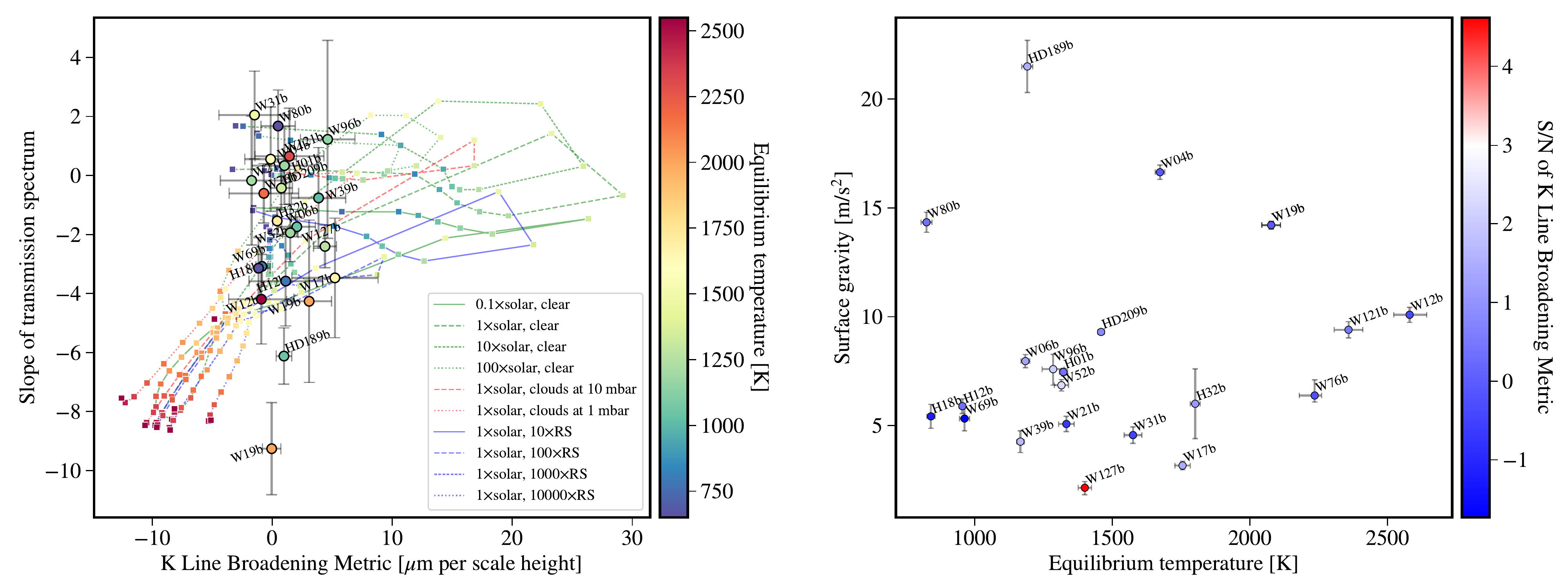}
\caption{{\it Left panels:} distribution of hot Jupiters characterized by low-resolution optical transmission spectroscopy on the plane of spectral slope versus line broadening metric. The first and second rows show the metric for Na and K, respectively. The colored lines show the corresponding values measured in the fiducial models, which connect models of different temperatures in the same group, that is, different metallicities, different cloud conditions, or different Rayleigh scattering (RS) enhancements. {\it Right panels: } distribution of the same hot Jupiters on the plane of planetary surface gravity versus equilibrium temperature. They are color coded by the signal-to-noise ratio (S/N) of the measured line broadening metric. The white part of the colorbar is centered at $S/N=3$. \label{fig:LBM}}
\end{figure*}

\begin{table*}
     \centering
     \caption{Measured line broadening metric and spectral slope for 24 hot Jupiters with low-resolution optical transmission spectroscopy.}
     \label{tab:ew_param}
     \begin{tabular}{cccccccccc}
     \hline\hline\noalign{\smallskip}
     Planet & $T_\mathrm{eq}$ \tablefootmark{(a)} & $g_\mathrm{p}$ \tablefootmark{(a)} & $R_\star$ \tablefootmark{(a)} & LBM(Na) & LBM(K) & $\alpha$ \tablefootmark{(b)} & $H$ \tablefootmark{(c)} & $\mathcal{T}$ \tablefootmark{(d)} & Reference\\\noalign{\smallskip}
            & [K]             & [m\,s$^{-1}$]  & [$R_\sun$] &  [$\mu$m/$H$] &  [$\mu$m/$H$] &          &  [$R_\star$]   &  [ppm]         &\\\noalign{\smallskip}
     \hline\noalign{\smallskip}
HAT-P-1b   & $1322 ^{+ 14 }_{- 15 }$ & $ 7.46 ^{+0.17 }_{-0.17 }$ &  1.174 & $ 2.21 \pm 2.55$ & $ 1.02 \pm 2.69$ & $ 0.3 \pm 1.3$ & 0.0008 &   180 & 1\\ \noalign{\smallskip}
HAT-P-12b  & $ 955 ^{+ 11 }_{- 11 }$ & $ 5.89 ^{+0.34 }_{-0.34 }$ &  0.679 & $-1.16 \pm 1.85$ & $ 1.12 \pm 3.06$ & $-3.6 \pm 1.5$ & 0.0012 &   343 & 1, 2\\ \noalign{\smallskip}
HAT-P-18b  & $ 841 ^{+ 15 }_{- 15 }$ & $ 5.42 ^{+0.55 }_{-0.55 }$ &  0.717 & $-1.07 \pm 1.17$ & $-1.14 \pm 0.96$ & $-3.1 \pm 1.0$ & 0.0011 &   303 & 3\\ \noalign{\smallskip}
HAT-P-32b  & $1801 ^{+ 18 }_{- 18 }$ & $ 6.00 ^{+1.60 }_{-1.60 }$ &  1.225 & $ 0.25 \pm 0.39$ & $ 0.41 \pm 0.37$ & $-1.5 \pm 0.5$ & 0.0013 &   383 & 4, 5, 6, 7\\ \noalign{\smallskip}
HAT-P-41b  & $1941 ^{+ 38 }_{- 38 }$ & $ 6.90 ^{+1.00 }_{-1.00 }$ &  1.683 & $ 3.65 \pm 2.10$ & -- & $-2.4 \pm 2.6$ & 0.0009 &   178 & 8\\ \noalign{\smallskip}
HD 189733b & $1191 ^{+ 20 }_{- 20 }$ & $21.50 ^{+1.20 }_{-1.20 }$ &  0.752 & $-0.47 \pm 0.75$ & $ 0.98 \pm 0.65$ & $-6.1 \pm 1.0$ & 0.0004 &   120 & 1\\ \noalign{\smallskip}
HD 209458b & $1459 ^{+ 12 }_{- 12 }$ & $ 9.30 ^{+0.08 }_{-0.08 }$ &  1.162 & $ 5.80 \pm 1.12$ & $ 0.76 \pm 0.92$ & $-0.4 \pm 0.7$ & 0.0007 &   170 & 1\\ \noalign{\smallskip}
WASP-4b    & $1673 ^{+ 17 }_{- 17 }$ & $16.64 ^{+0.33 }_{-0.33 }$ &  0.910 & $ 1.72 \pm 0.60$ & $-0.13 \pm 2.16$ & $ 0.6 \pm 1.8$ & 0.0006 &   176 & 9\\ \noalign{\smallskip}
WASP-6b    & $1184 ^{+ 16 }_{- 16 }$ & $ 7.96 ^{+0.30 }_{-0.30 }$ &  0.864 & $-1.85 \pm 1.53$ & $ 2.05 \pm 1.55$ & $-1.7 \pm 0.3$ & 0.0009 &   260 & 1\\ \noalign{\smallskip}
WASP-12b   & $2580 ^{+ 58 }_{- 62 }$ & $10.09 ^{+0.35 }_{-0.34 }$ &  1.657 & $-1.62 \pm 3.63$ & $-0.91 \pm 2.72$ & $-4.2 \pm 1.5$ & 0.0008 &   188 & 1\\ \noalign{\smallskip}
WASP-17b   & $1755 ^{+ 28 }_{- 28 }$ & $ 3.16 ^{+0.20 }_{-0.20 }$ &  1.583 & $-5.77 \pm 4.02$ & $ 5.22 \pm 3.60$ & $-3.5 \pm 2.0$ & 0.0018 &   454 & 1\\ \noalign{\smallskip}
WASP-19b   & $2077 ^{+ 34 }_{- 34 }$ & $14.21 ^{+0.18 }_{-0.18 }$ &  1.018 & $-2.69 \pm 1.25$ & $-0.04 \pm 0.77$ & $-9.3 \pm 1.6$ & 0.0007 &   211 & 10 \tablefootmark{(e)}\\ \noalign{\smallskip}
WASP-19b   & $2077 ^{+ 34 }_{- 34 }$ & $14.21 ^{+0.18 }_{-0.18 }$ &  1.018 & $-1.69 \pm 1.86$ & $ 3.08 \pm 1.87$ & $-4.3 \pm 2.8$ & 0.0007 &   211 & 11 \tablefootmark{(e)}\\ \noalign{\smallskip}
WASP-21b   & $1333 ^{+ 28 }_{- 28 }$ & $ 5.07 ^{+0.35 }_{-0.35 }$ &  1.136 & $24.04 \pm 3.10$ & $-1.72 \pm 2.61$ & $-0.2 \pm 2.2$ & 0.0012 &   251 & 12\\ \noalign{\smallskip}
WASP-29b   & $ 970 ^{+ 32 }_{- 31 }$ & $10.00 ^{+1.40 }_{-1.40 }$ &  0.808 & $ 3.30 \pm 7.05$ & -- & $-1.6 \pm 2.1$ & 0.0006 &   122 & 13\\ \noalign{\smallskip}
WASP-31b   & $1575 ^{+ 32 }_{- 32 }$ & $ 4.56 ^{+0.38 }_{-0.38 }$ &  1.252 & $-3.12 \pm 1.77$ & $-1.47 \pm 2.97$ & $ 2.0 \pm 1.5$ & 0.0014 &   362 & 1\\ \noalign{\smallskip}
WASP-39b   & $1166 ^{+ 14 }_{- 14 }$ & $ 4.26 ^{+0.50 }_{-0.50 }$ &  0.939 & $ 4.88 \pm 1.55$ & $ 3.86 \pm 2.26$ & $-0.8 \pm 1.7$ & 0.0015 &   421 & 1\\ \noalign{\smallskip}
WASP-52b   & $1315 ^{+ 26 }_{- 26 }$ & $ 6.85 ^{+0.26 }_{-0.26 }$ &  0.786 & $-1.70 \pm 0.73$ & $ 1.51 \pm 0.65$ & $-1.9 \pm 1.0$ & 0.0013 &   413 & 14, 15\\ \noalign{\smallskip}
WASP-69b   & $ 963 ^{+ 18 }_{- 18 }$ & $ 5.32 ^{+0.56 }_{-0.56 }$ &  0.813 & $-0.40 \pm 1.38$ & $-0.86 \pm 0.50$ & $-3.1 \pm 0.7$ & 0.0011 &   307 & 16\\ \noalign{\smallskip}
WASP-76b   & $2235 ^{+ 56 }_{- 25 }$ & $ 6.38 ^{+0.30 }_{-0.72 }$ &  1.716 & $ 1.26 \pm 1.07$ & $-0.70 \pm 2.90$ & $-0.6 \pm 1.0$ & 0.0011 &   238 & 17\\ \noalign{\smallskip}
WASP-80b   & $ 825 ^{+ 20 }_{- 20 }$ & $14.34 ^{+0.46 }_{-0.46 }$ &  0.593 & $ 0.66 \pm 1.88$ & $ 0.49 \pm 1.42$ & $ 1.7 \pm 1.2$ & 0.0005 &   171 & 18\\ \noalign{\smallskip}
WASP-96b   & $1285 ^{+ 40 }_{- 40 }$ & $ 7.59 ^{+0.73 }_{-0.68 }$ &  1.050 & $ 9.83 \pm 2.09$ & $ 4.62 \pm 2.27$ & $ 1.2 \pm 3.4$ & 0.0008 &   195 & 19\\ \noalign{\smallskip}
WASP-121b  & $2358 ^{+ 52 }_{- 52 }$ & $ 9.40 ^{+0.37 }_{-0.37 }$ &  1.458 & $ 1.57 \pm 2.07$ & $ 1.43 \pm 2.87$ & $ 0.7 \pm 1.6$ & 0.0009 &   233 & 20\\ \noalign{\smallskip}
WASP-127b  & $1400 ^{+ 24 }_{- 24 }$ & $ 2.14 ^{+0.32 }_{-0.28 }$ &  1.390 & $ 2.85 \pm 0.87$ & $ 4.40 \pm 0.95$ & $-2.4 \pm 0.7$ & 0.0024 &   492 & 21\\ \noalign{\smallskip}
XO-2b      & $1328 ^{+ 17 }_{- 28 }$ & $14.13 ^{+1.01 }_{-0.94 }$ &  0.998 & $ 6.29 \pm 1.48$ & -- & $ 0.2 \pm 0.7$ & 0.0005 &   102 & 22\\ \noalign{\smallskip}
 \hline\noalign{\smallskip}
    \end{tabular}
    \tablefoot{
      \tablefoottext{a}{Equilibrium temperature $T_\mathrm{eq}$, surface gravity $g_\mathrm{p}$, and stellar radius $R_\star$ are taken from TEPCat \citep{2011MNRAS.417.2166S}.}
      \tablefoottext{b}{Scattering slope of the transmission spectrum between 510--900~nm, excluding two 80~nm bands centered at Na and K, respectively.}
      \tablefoottext{c}{$H=k_\mathrm{B}T_\mathrm{eq}/\mu g_\mathrm{p}$ is the atmospheric scale height, where $\mu=2.3$~g\,mol$^{-1}$ is assumed.}
      \tablefoottext{d}{$\mathcal{T}=2HR_\mathrm{p}/R_\star^2$ is the transmission signal per scale height.}
      \tablefoottext{e}{Due to discrepant results, the measurements for WASP-19b are listed individually, while weighted average values are presented for the other planets when multiple observations are available.}
    }
    \tablebib{
(1) \citet{2016Natur.529...59S};  
(2) \citet{2020AJ....159..234W};  
(3) \citet{2017MNRAS.468.3907K};  
(4) \citet{2013MNRAS.436.2974G};  
(5) \citet{2016A&A...590A.100M};  
(6) \citet{2016A&A...594A..65N}; 
(7) \citet{2020AJ....160...51A}; 
(8) \citet{2020AJ....159..204W};  
(9) \citet{2017AJ....154...95H};  
(10) \citet{2017Natur.549..238S};  
(11) \citet{2019MNRAS.482.2065E};  
(12) This work; 
(13) \citet{2013MNRAS.428.3680G};  
(14) \citet{2017A&A...600L..11C};  
(15) \citet{2018AJ....156..298A};  
(16) \citet{2020arXiv200702741M}; 
(17) \citet{2020arXiv200502568F};  
(18) \citet{2018A&A...609A..33P};  
(19) \citet{2018Natur.557..526N};  
(20) \citet{2018AJ....156..283E};  
(21) \citet{2018A&A...616A.145C};  
(22) \citet{2019AJ....157...21P}.
}
\end{table*}

To put the detected Na line broadening of WASP-21b in a general context of the hot Jupiter population, we define a line broadening metric (LBM) as follows:
\begin{equation}
\mathrm{LBM}=\frac{1}{\mathcal{T}}\sum^{+40~\mathrm{nm}}_{\lambda=-40~\mathrm{nm}}\frac{\mathcal{D}_\lambda-\mathcal{D}_0}{\mathcal{D}_0}d\lambda,
\end{equation}
where $d\lambda$ is the band width of a given passband in the low-resolution transmission spectrum, $\mathcal{T}=2HR_\mathrm{p}/R_\star^2$ is the expected transmission signal per scale height $H$, and $\mathcal{D}_0=(R_\mathrm{p}^2/R_\star^2)_0$ is the linearly interpolated continuum based on two 100~nm bands bracketing the 80~nm band centered at the Na line (or the K line). These three bands could be composed of different number of passbands, depending on how the literature studies were reporting their low-resolution transmission spectra. This metric becomes the Na or K equivalent width if a pressure-broadened Na or K line is detected. We measure LBM for all the hot Jupiters that have been studied by low-resolution transmission spectroscopy and that have sufficient wavelength coverage and passband resolution. In addition to LBM, we also calculate the local spectral slope $\alpha$ of these transmission spectra within the wavelength range of 510--900~nm, excluding two 80~nm bands centered at Na and K, by fitting a linear line in the ($\ln\lambda$, $R_\mathrm{p}/R_\star$) space, that is,
\begin{equation}
\alpha=\frac{1}{H}\frac{dR_\mathrm{p}}{d\ln\lambda}.
\end{equation}
The measured LBM for both Na and K and the local spectral slope $\alpha$ are given in Table \ref{tab:ew_param}.

To compare this observational metric to theoretical predictions, we then create a set of fiducial isothermal model transmission spectra using the \texttt{Exo-Transmit} code \citep{2017PASP..129d4402K}, adopting WASP-21b's bulk parameters and covering temperatures from 650~K to 2650~K. The adopted gas opacities are: Na, K, TiO, VO, H$_2$O, CH$_4$, CO, CO$_2$. The other considerations include metallicities of 0.1$\times$, 1$\times$, 10$\times$, and 100$\times$solar, clouds at 10~mbar, 1~mbar or cloud-free, Rayleigh scattering enhanced by 1$\times$, 10$\times$, 1000$\times$, and 10000$\times$. We calculate LBM and spectral slope $\alpha$ for these models in the same way as the transmission spectrum data. 

Figure~\ref{fig:LBM} presents the distribution of 23 hot Jupiters collected from literature studies, along with WASP-21b. The left panels of Fig.~\ref{fig:LBM} show the distribution on the plane of spectral slope versus LBM. Most hot Jupiters have the LBM values consistent with zero, indicative of no excess absorption within the 80~nm band centered at Na or K. Corresponding spectral slope varies from $-$9 to $+$2, which do not necessarily represent scattering features alone, because it is derived locally where absorption from molecules such as TiO and VO could bias it from a pure scattering slope. WASP-21b stands out with the largest Na LBM value, followed by WASP-96b \citep{2018Natur.557..526N} and XO-2b \citep{2019AJ....157...21P}. All these three hot Jupiters have shown clear pressure broadening at the Na line in low-resolution transmission spectra. This is consistent with the predictions of the fiducial models: the upper right corner is the location of hot Jupiters with clear atmospheres exhibiting significant alkali line broadening. On the other hand, only WASP-127b \citep{2018A&A...616A.145C} shows a significant K LBM value, while the others have large uncertainties. The regions where TiO/VO-dominated atmospheres are predicted to locate, are free of any measurements. This is a natural result of the lack of TiO/VO detections in low-resolution optical transmission spectroscopy. The presence of clouds at higher altitudes or hazes introducing enhanced Rayleigh scattering would move the measurements towards zero LBM value.

The right panels of Fig.~\ref{fig:LBM} show the distribution of hot Jupiters on the plane of surface gravity versus equilibrium temperature. They are assigned with colors according to the S/N of the measured LBM. The colormap is adjusted so that the white color is centered at $S/N=3$. Consequently, the hot Jupiters with $S/N>3$ would appear reddish. It is striking that the hot Jupiters with high significance of LBM have similar equilibrium temperatures, between $\sim$1200~K and $\sim$1500~K, and spanning a wide range of surface gravity values. WASP-21b is the one with the most significant LBM value, followed by WASP-96b, XO-2b, HD 209458b, WASP-127b, and WASP-39b. Indeed, in addition to WASP-21b, all other five hot Jupiters have been reported to have relatively clear atmospheres \citep{2018Natur.557..526N,2019AJ....157...21P,2017MNRAS.469.1979M,2018A&A...616A.145C,2018AJ....155...29W}. Therefore, LBM can serve as a good indicator of pressure broadening of alkali lines, providing a new path to quantitatively compare the atmospheres of different hot Jupiters.

\section{Conclusions}
\label{sec:conclusions}

We observed one transit of the Saturn-mass planet \object{WASP-21b} with the low-resolution spectrograph OSIRIS at the 10.4~m GTC. We derived a transmission spectrum composed of 37 spectral bins with a uniform width of 10~nm. The most prominent spectral signature is a broad profile centered at the Na doublet, which is likely associated with the pressure broadening. The transmission spectrum shows a tentative evidence of excess absorption at the K D$_1$ line. We performed a spectral retrieval analysis on this transmission spectrum and reported the detection of Na at a confidence level of $>$3.5-$\sigma$. While a fiducial model leads to a Na detection at 3.5-$\sigma$ significance, a simplified model provides a detection at 4.9-$\sigma$ significance. The current data quality is not sufficient to constrain the chemical abundance and temperature structure precisely, for which high-precision follow-up observations are required.

We also observed one transit of WASP-21b with the high-resolution spectrograph HARPS-N at the 3.58~m TNG, and collected the archival data of another two transits observed by HARPS at the ESO 3.6~m telescope. The measured radial velocities exhibit a Rossiter-McLaughlin anomaly consistent with an aligned planetary orbit to stellar spin axis. We performed high-resolution transmission spectroscopy analysis and detected an excess absorption at the Na doublet. The resolved Na doublet shows a radial velocity shift during the transit that is consistent with the planet orbital motion, confirming its planetary origin. The Na doublet also exhibits a tentative net blueshift that might hint a day-to-night wind. The data do not reveal any significant excess absorption at other atomic species. 

With the data at hand, and comparing with literature results from other hot Jupiter observations, we proposed the line broadening metric (LBM) to quantify the excess absorption around the Na or K line in low-resolution optical transmission spectra. We measured the LBM values for 24 hot Jupiters that have been previously characterized by low-resolution optical transmission spectroscopy, and found that the hot Jupiters with high LBM values are likely to exhibit pressure-broadened Na line profiles. This slope-LBM diagram is the exoplanet version of color-color diagram, which can quantitatively distinguish the relatively clear atmospheres from others and thus help prioritize targets for observing campaigns. The metric of current hot Jupiter collection reveals that relatively clear atmospheres appear at equilibrium temperatures of 1200--1500~K. WASP-21b has the largest LBM value among the current collection. Together with the fact that Na is clearly detected in both low- and high-resolution transmission spectrum, WASP-21b is likely to have a relatively clear atmosphere, and thus will be an extremely interesting target for {\it James Webb} space telescope.

\begin{acknowledgements}
    G.\,C. acknowledges the support by the B-type Strategic Priority Program of the Chinese Academy of Sciences (Grant No.\,XDB41000000), the Natural Science Foundation of Jiangsu Province (Grant No.\,BK20190110), and the Minor Planet Foundation of the Purple Mountain Observatory. 
    This work is partly financed by the Spanish Ministry of Economics and Competitiveness through grant ESP2013-48391-C4-2-R. 
    This work is based on observations made with the Gran Telescopio Canarias (GTC), installed at the Spanish Observatorio del Roque de los Muchachos of the Instituto de Astrof\'{i}sica de Canarias, in the island of La Palma. 
    This work is based on observations made with the Italian Telescopio Nazionale {\it Galileo} (TNG) operated on the island of La Palma by the Fundaci\'{o}n Galileo Galilei of the INAF (Istituto Nazionale di Astrofis\'{i}ca) at the Spanish Observatorio del Roque de los Muchachos of the Instituto de Astrofisica de Canarias. 
    This work is based on observations made with the Telescopio Nazionale {\it Galileo} (TNG) under Director's Discretionary Time of Spain's Instituto de Astrof\'{i}sica de Canarias.
    This work has made use of Matplotlib \citep{2007CSE.....9...90H} and the VizieR catalog access tool, CDS, Strasbourg, France \citep{2000A&AS..143...23O}. 
    The authors thank the anonymous referee for their constructive comments on the manuscript.
\end{acknowledgements}

\bibliographystyle{aa} 
\bibliography{ref_db.bib} 


\begin{appendix}
\onecolumn

\section{Additional figures for the GTC observation.}
\label{sec:app_lowres}

\begin{figure*}[h!]
\centering
\includegraphics[width=1.0\linewidth]{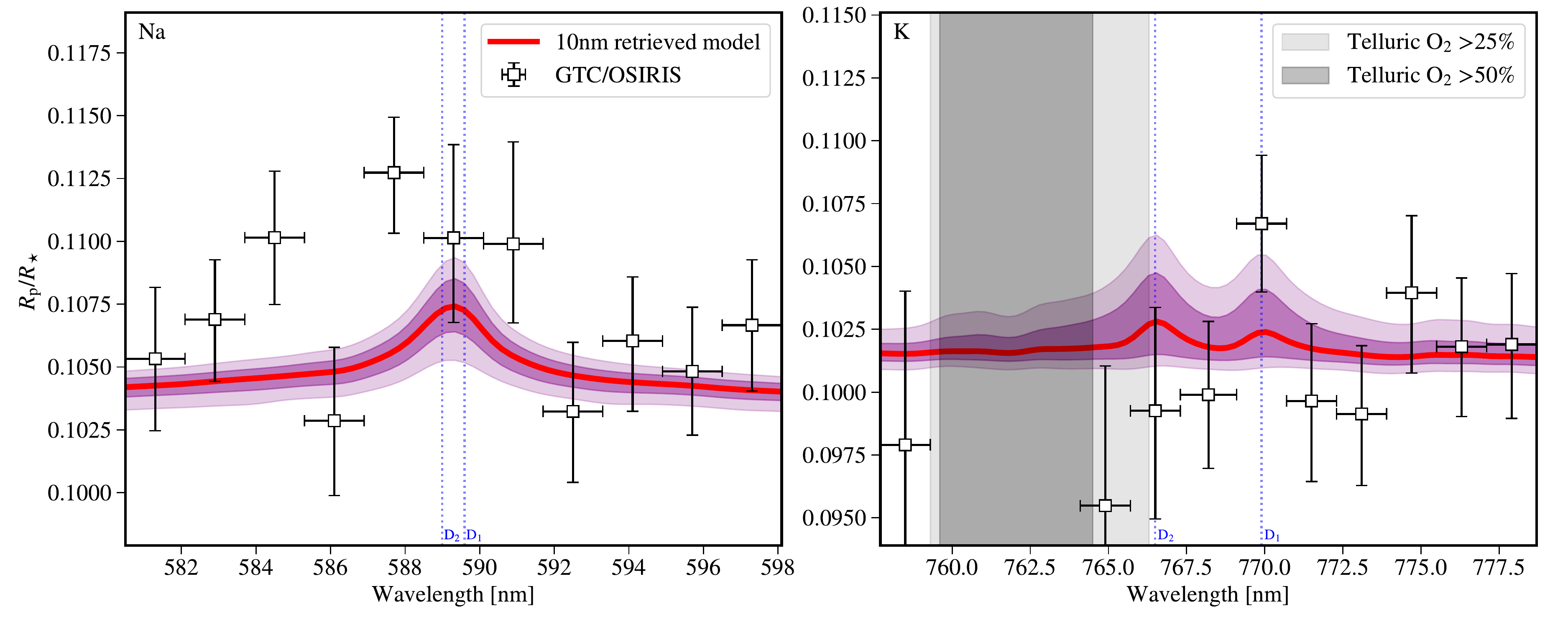}
\caption{GTC/OSIRIS 16~$\AA$ bin transmission spectrum, zoomed at the Na ({\it left}) and K ({\it right}) doublets. The red line along with the purple shaded regions show the best model and its 1-$\sigma$/2-$\sigma$ confidence levels retrieved from the 10~nm bin transmission spectrum. While it is too noisy to claim detections of the Na and K line cores, the data indeed hint possible excess absorption at the Na doublet and the K D$_1$ line. The K D$_2$ line is located in the telluric oxygen-A band.}
\label{fig:GTC_NaK}
\end{figure*}

\section{Additional figures for the HARPS-N and HARPS observations.}
\label{sec:app_hires}

\begin{figure*}[h!]
\centering
\includegraphics[width=1.0\linewidth]{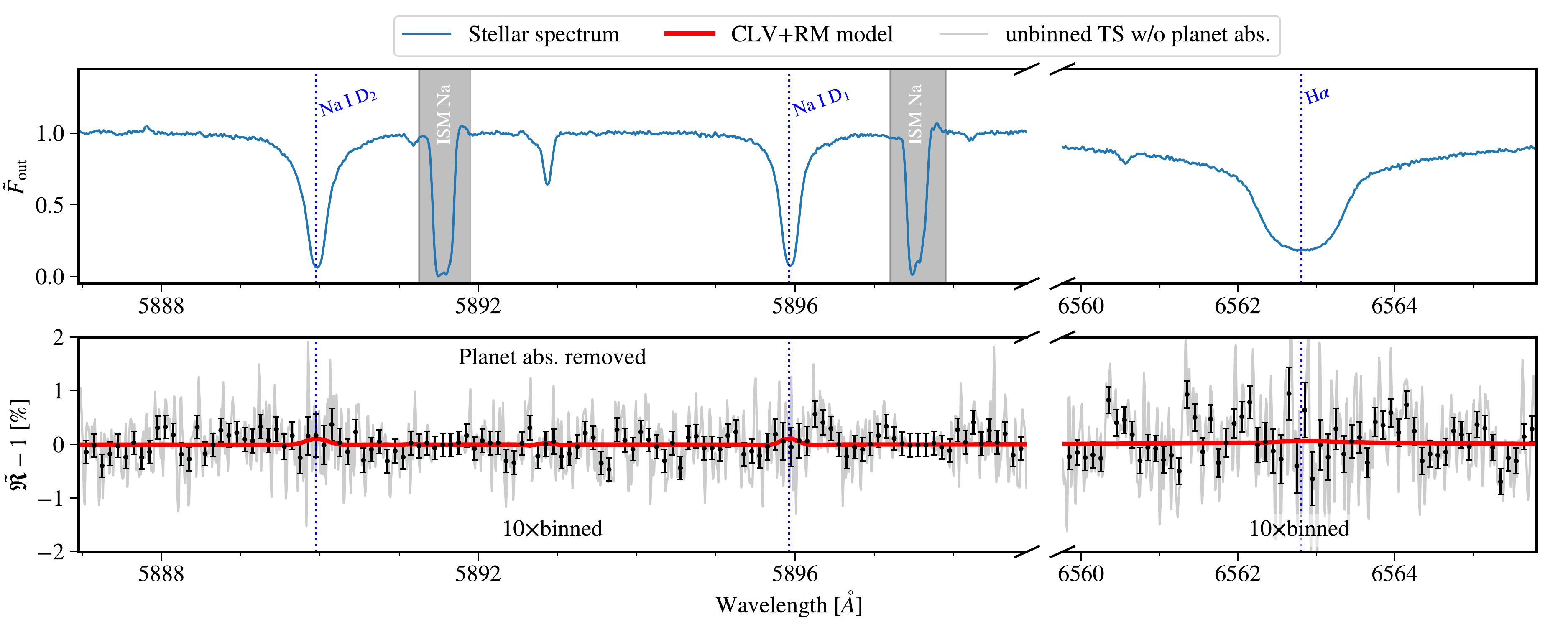}
\caption{{\it Top panel}: out-of-transit master stellar spectrum at the Na doublet and H$\alpha$ lines. The strong interstellar Na absorption, indicated by the gray shadow, can be noticed on the red side of the stellar Na absorption. The wavelength is in the stellar rest frame. {\it Bottom panel}: transmission spectrum without the center-to-limb variation (CLV) and Rossiter-McLaughlin (RM) correction. The best-fit planet absorption presented in Fig.~\ref{fig:hires_transpec} has been removed. The red line shows the combined CLV and RM model.}
\label{fig:NaHa_CLV_RM}
\end{figure*}

\begin{figure*}[h!]
\centering
\includegraphics[width=1.0\linewidth]{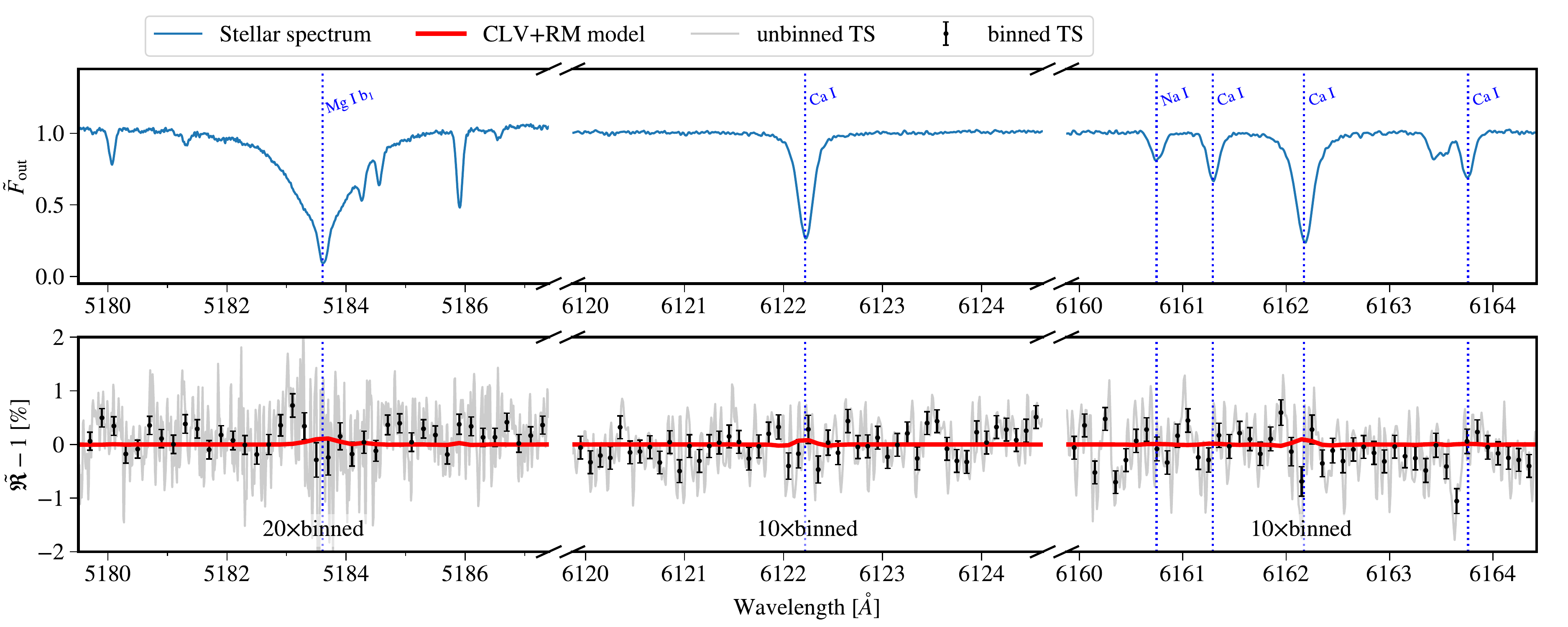}
\caption{{\it Top panel}: out-of-transit master stellar spectrum at several stellar activity indicator lines. {\it Bottom panel}: transmission spectrum without the center-to-limb variation (CLV) and Rossiter-McLaughlin (RM) correction. Given the non-detection of excess absorption, no removal of planet absorption is performed. The red line shows the combined CLV and RM model.}
\label{fig:activity_indicators}
\end{figure*}

\end{appendix}

\end{document}